\documentclass[twoside]{article}
\usepackage{graphicx}
\usepackage[latin1]{inputenc}
\usepackage[a4paper,tmargin=3cm, bmargin=3cm, lmargin=3cm, rmargin=3cm]{geometry}
\usepackage{amsmath,amssymb,amsthm,amscd,graphicx,color,accents,float}

\usepackage{mathrsfs,natbib}
\usepackage{tocbibind}

\usepackage{algorithm}
\usepackage[noend]{algpseudocode}

\usepackage[colorlinks=true,linkcolor=black,citecolor=black,urlcolor=black]{hyperref} 

\numberwithin{equation}{section}

\theoremstyle{plain}
\newtheorem{theorem}{Theorem}[section]
\newtheorem{proposition}[theorem]{Proposition}
\newtheorem{lemma}[theorem]{Lemma}
\newtheorem{corollary}[theorem]{Corollary}
\newtheorem{remark}[theorem]{Remark}

\DeclareMathOperator*{\argmin}{arg\,min}

\def\tr{{\rm{tr}\,}}

\def\bD{{\boldsymbol{D}}}

\def\bK{{\boldsymbol{K}}}

\def\bU{{\boldsymbol{U}}}
\def\bX{{\boldsymbol{X}}}
\def\bY{{\boldsymbol{Y}}}
\def\bZ{{\boldsymbol{Z}}}
\def\mX{{\mathcal{X}}}
\def\mY{{\mathcal{Y}}}
\def\mZ{{\mathcal{Z}}}

\def\P{{\mathbb P}}

\def\V{{\mathcal V}}

\def\E{\mathbb{E}}

\def\R{\mathbb{R}}

\def\BState{\State\hskip-\ALG@thistlm}

\def\spx{\mathcal{X}}

\usepackage{sectsty,titlesec}

\titlelabel{\thetitle.~}
\sectionfont{\normalsize\bfseries\centering}
\subsectionfont{\normalsize\bfseries\centering}

\begin{document}

\begin{center}

\textbf{\large A GENERALIZED DISTANCE COVARIANCE FRAMEWORK\\[11pt] FOR GENOME-WIDE ASSOCIATION STUDIES}



\vspace{0.15cm}

Dominic Edelmann$^{1}$, Fernando Castro-Prado$^{2}$ and Jelle J. Goeman$^{3}$ 
\end{center}

\vspace{0.06cm}

$^{1}$German Cancer Research Center (DKFZ), Heidelberg, Germany.

$^{2}$University and Health Research Institute (USC and IDIS), Santiago de Compostela, Spain.

$^{3}$Leiden University Medical Center (LUMC), Leiden, the Netherlands.

\section*{ABSTRACT} 
\parskip.5cm
\begin{quotation}

When testing for the association of a single SNP with a phenotypic response, one usually considers an additive genetic model, assuming that the mean of of the response for the heterozygous state is the average of the means for the two homozygous states. However, this simplification often does not hold. In this paper, we present a novel framework for testing the association of a single SNP and a phenotype. Different from the predominant standard approach, our methodology is guaranteed to detect all dependencies expressed by classical genetic association models. The asymptotic distribution under mild regularity assumptions is derived. Moreover, the  finite sample distribution under Gaussianity is provided in which the exact p-value can be efficiently evaluated via the classical Appell hypergeometric series. Both results are extended to a regression-type setting with nuisance covariates, enabling hypotheses testing in a wide range of scenarios. A connection of our approach to score tests is explored, leading to intuitive interpretations as locally most powerful tests. A simulation study demonstrates the computational efficiency and excellent statistical performance of the proposed methodology. A real data example is provided.


\end{quotation}\parskip-0cm

\vspace{0.4cm}

\textbf{Keywords}: Genome-wide analysis studies; Genomic data; Single nucleotide polymorphisms; Partially dominant models; Distance covariance; Global tests; Hilbert--Schmidt independence criterion; Feature maps.

\vspace{0.4cm}

\section{INTRODUCTION}


Since the days of Gregor Mendel, a key motivation of genetic research has been to understand the link between genotype and phenotype \citep{Zschocke,Brandes}. With the beginning of the Human Genome Project in the 1990s, one of the major goals has been to sequence the DNA of large cohorts of individuals to unravel the molecular causes of human trait variation \citep{Lander1996}. Over the last 15 years, genome-wide association studies (GWASs) have evolved from a promising idea to a reality that has revolutionized the way research in human trait genetics is conducted \citep{15y}. 



GWAS involves testing genetic variants across the genomes of many individuals to identify genotype-phenotype associations. Thus, the response variable corresponds to a phenotypic characteristic of interest; in this work we will restrict ourselves to continuous response variables (e.g., physical measures such as height, concentration of certain molecules in the blood, cardiological parameters). 

Despite the diversity of existing technologies to analyze the human genome, GWAS databases typically focus on \emph{single-nucleotide polymorphisms} \setcitestyle{square}(SNPs; \citet{Tam}),\setcitestyle{round} which are the most simple and common form of genetic variation among humans. Each SNP represents a change in one of the 3 billion letters that form the ``book of life'', that is, the alternation between the most frequent nucleotide in that specific genomic position in the general human population, and another nucleotide that can be observed in a minority of people (and therefore called \emph{minor allele}). For instance, let us assume that a certain SNP has $A_1$ as the wild type (the most frequent nucleotide at this position), and $A_2$ as the ``mutation'' (the less frequent nucleotide). Then each individual will have one of the three following genotypes in their (diploid) genome:$$\{A_1 A_1,A_1 A_2, A_2 A_2\}.$$
The aim is to find SNPs that are associated with the trait of interest.

Due to particularities of the biological background (most pairs of SNPs are approximately independent), it is often meaningful to test for the association of each single SNP with the response variable. For this purpose, it is standard to apply an \textit{additive} model \citep{Brandes}, coding the three possible states by counting the number of minor alleles, i.e. in our example,$$0:=A_1 A_1;\;\,\;1:=A_1 A_2,\;\;\;2:= A_2 A_2,$$
and subsequently treating this variable as continuous. If no additional nuisance covariates are considered, this model implies that the mean of the heterozygous state ($A_1 A_2$) is the average of the means for the two homozygous states ($A_1 A_1$ and $A_2 A_2$). However, it has been shown that these approaches often lead to suboptimal results, which calls for the development of new methods to infer genotype-phenotype association \citep{SIM}.

In this work, we present a novel approach for testing the association with a quantitative response variable based on the concept of generalized distance covariance \citep{SRB,szekely2009brownian,Sejdinovic,DJ}. We consider defining different metrics on the genotype space (in our example $\{A_1 A_1 ,A_1 A_2,A_2 A_2\}$), each of these differences leading to a different distance covariance and hence test statistic. Fixing the distance between the homozygous states and the heterozygous state at $1$ (in our example $d(A_1 A_1,A_1 A_2)=d(A_2 A_2,A_1 A_2)=1$), metrics can then be expressed by a single parameter, the distance between the two homozygous states (in our example $b := d(A_1 A_1,A_2 A_2)$). We show that, different from standard methods, our method is guaranteed to detect any of the classical genetic association models in \citet{GI} (cf. Section \ref{Models:SNP}) if $b$ is chosen in the open interval $(0,4)$. To enable efficient testing, both the asymptotic distribution of the test statistic and the finite-sample distribution under Gaussianity are derived. In the latter case, the exact p-value can be efficiently evaluated via the classical Appell hypergeometric series \citep{Appell}, which is a clear advantage compared to permutation procedures (that are too slow to calculate the small p-values up to the desired precision) or Gamma approximation methods (that are not guaranteed to lead to valid tests).

A connection of our testing procedures with locally most powerful tests in a Gaussian regression setting \citep{goeman2006testing,goeman2011testing} is derived, providing valuable insights from a biological perspective. We show that, for each choice of $b$, the corresponding test can be interpreted as the locally most powerful test for which certain fractions of SNP-phenotype associations follow an additive, dominant, recessive or purely heterozygous model, respectively. Alternatively, each test can be interpreted as the locally most powerful test for which the heterozygous effect parameter $h$ follows a certain distribution, determined by $b$. All testing procedures are extended to be used with nuisance covariates and imputed data, which enables them to be applied in virtually all standard GWAS setting. Simulations demonstrate an excellent performance of the proposed testing methodology. Notably, the test for $b=3$ shows only a small disadvantage in power against the additive model in some scenarios, but features clear power gains in multiple others.


Several other works have proposed distance covariance methods to analyze GWAS data. \citet{Hua} proposed distance as means to analyze the association of SNPs with neuroimaging data using a Gamma approximation for their testing procedure. \citet{Fischer:kernel} use a distance covariance approach for testing association of genes and multiple phenotypes in case-parent trios. Apart from genome-wide testing, distance covariance has been applied for feature screening \citep{Li:SIS} in regularized regression on genomic data \citet{Carlsen,Rongling} and various purposes in other types of omics data \citep{Guo,SysBio,ghanbari2019}.

The rest of the article is structured as follows. Section~\ref{Background} introduces some additional genetic concepts, as well as concept of generalized distance covariance. In Section~\ref{sec:testing} we introduce the testing approach and prove the asymptotic and finite-sample distribution; we also provide some details on the numerical approximation of the p-values in practice. Section~\ref{GT} provides local optimality properties of each of the proposed statistic, leading to important insights which choice of $b$ is appropriate in applications. The theory is extended to involve nuisance covariates in Section~\ref{Covariates}. Extensions to imputed data and multi-allelic SNPs are discussed in Section~\ref{sec:practical}. Detailed simulation studies in Section~\ref{sec:simu} demonstrate the computational efficiency and good statistical performance of our methods. A real data example is provided in Section~\ref{sec:rd}. We conclude with a general discussion and some final remarks in Section~\ref{Discu}. 

\section{BACKGROUND}\label{Background}

\subsection{Models for the association between SNPs and quantitative traits}\label{Models:SNP}


We now introduce some concepts of basic quantitative genetics modeling, which will allow us to interpret our testing framework and the results it offers from a biological perspective.

As previously mentioned, we will assume 
that SNPs are biallelic loci, i.e., they can present the major allele $A_1$ or the minor allele $A_2$, with the latter having a lower frequency in the population. With this notation, the three possible genotypes each individual can carry are: $A_1 A_1$ (major allele in homozygosity), $A_1 A_2$ (heterozygosity) and $A_2 A_2$ (minor allele in homozygosity). To be consistent with standard genetic notation, we will encode those three genotypes as the values of a random element $X$ with support $\{0,1,2\}$, which counts the occurrences of the second allele.

\begin{table}[H]
\begin{center}
\begin{tabular}{r|rrr}
& $X=0 \, (A_1 A_1)$ & $X=1 \, (A_1 A_2)$ & $X=2 \, (A_2 A_2)$ \\
\hline
 mean of $Y$ conditional to $X$ & $\mu_0$ & $\mu_1$ &  $\mu_2$\\
 standardized effect (for $\mu_0 \neq \mu_2)$ &$0$ & $h$  & $1$ \\
\end{tabular}
\caption{Association models between a SNP $X$ and an absolutely continuous quantitative trait $Y$.}
\label{Table:Models:SNP}
\end{center}
\end{table}

For studying different models between the state $X \in \{0,1,2\}$ of a certain SNP and an absolutely continuous response $Y \in \R$, let us define the conditional mean of $Y$ given $X$: $$\mu_j = \E [Y|X=j],$$ where $j \in \{0,1,2\}$. In classical quantitative genetics \citep[][Section 3.2]{GI}, the association between $X$ and $Y$ is represented as on Table~\ref{Table:Models:SNP}, where one is generally assuming that the means of the two homozygous states are different: $\mu_2 \neq\mu_0$. The standardized effect for each state $j\in\{0,1,2\}$ is hereby calculated as $\frac{\mu_j-\mu_0}{\mu_2-\mu_0}$.

The association models are then classified based on the biological interpretation of the value of the parameter $h:=\frac{\mu_1-\mu_0}{\mu_2-\mu_0}\in\mathbb R$, known as {\em heterozygous effect}:

\begin{itemize}
    \item $h=0$: {\it dominant-recessive} model; where $A_1$ is {\it dominant}, $A_2$ is {\it recessive}.
    \item $h=1$: {\it dominant-recessive} model; where $A_2$ is \textit{dominant}, $A_1$ is \textit{recessive}.
    \item $h \in (0,1)$: {\it partially dominant model} (or incomplete dominance model). A partially dominant model with $h=\tfrac12$ is called {\it additive} model.
    \item $h<0$ or $h>1$: {\it overdominant} model.
\end{itemize}

\begin{figure}
    \centering
    \includegraphics[width=0.7\textwidth]{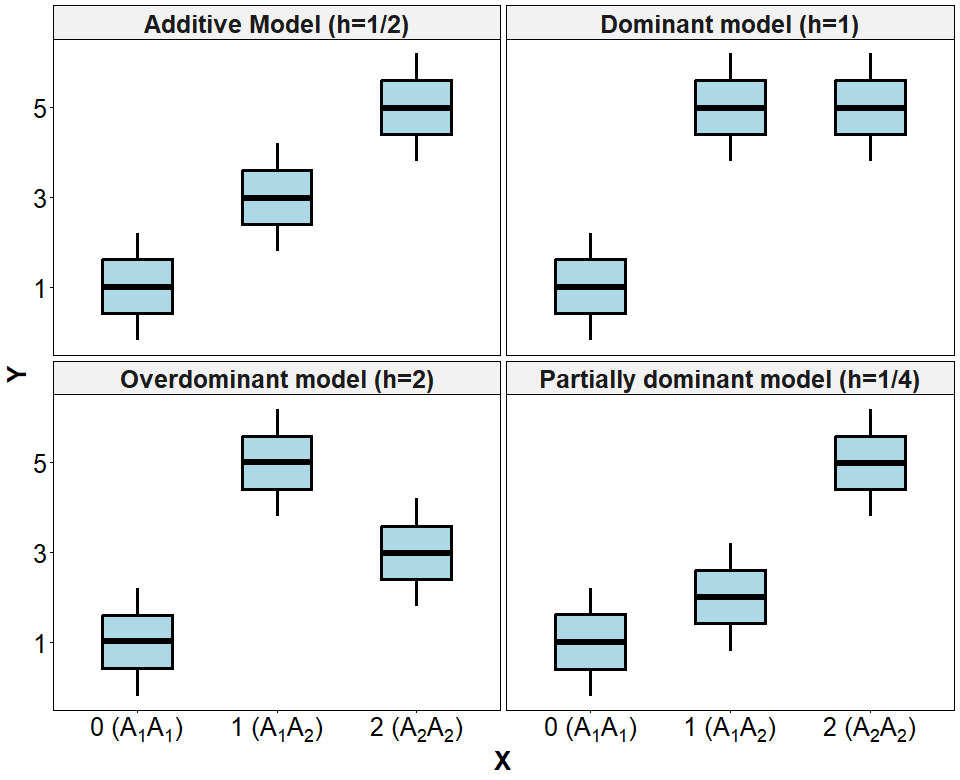}
    \caption{Some examples for classical genetic association models}.
    \label{fig:genmodels}
\end{figure}

In the course of this paper, we will also consider models for which $\mu_0 = \mu_2$ and $\mu_1 \neq \mu_0$, which we will refer to as {\em purely heterozygous}. Some examples for genetic association models are illustrated in Figure \ref{fig:genmodels}.

 For $h=0$ and $h=1$, $h$ corresponds to the dominance of (the phenotype of) $A_1$ and $A_2$ respectively, as current nomenclature of medical genetics indicates that dominance refers to the fact of observing the exact same phenotype of homozygosity also under heterozygosity \citep{Zschocke}. In this sense, the heterozygous effect $h$ can be (at least for dominant-recessive and partially dominant models) interpreted as the ``degree of dominance'' of $A_2$.

\subsection{Generalized distance covariance}\label{gdc}

Distance covariance and distance correlation \citep{SRB,szekely2009brownian} are measures of dependence that extend  classical product-moment covariance and Pearson's correlation. Given random variables $X \in \R^p$ and $Y \in \R^q$ with finite first moments, their distance covariance can be expressed as:
	\begin{equation}
	\label{eq:dcovalt}
	\V^2(X,Y) = \E\Big [\|X -X'\| \left\{ \|Y -Y' \| - \|Y-Y'' \|  
 - \|Y'-Y''\|+\|Y''-Y'''\| \right\} \Big], 
	\end{equation}
where primed letters denote i.i.d.~copies of $(X,Y)$ and $\|\cdot\|$ is the Euclidean norm.

In this work, we consider {\it generalized distance covariance} (GDC) in the terminology of \citet{Sejdinovic}, with the extension on premetrics described in \cite{DJ}.

Given a set $\mZ\neq\emptyset$, we say that a function $\rho: \mZ \times \mZ \to [0,\infty)$ is a premetric if it is symmetric in its arguments and satisfies $\rho(z,z)=0$ for all $z\in\mZ$. Then $(\mZ,\rho)$ is called a premetric space.

Moreover, we say that a premetric space has negative type that condition whenever  $\sqrt{\rho}$ satisfies the triangle inequality (see \cite[p. 2266]{Sejdinovic} for a motivation of this property and more details).
Now let $(\mX, \rho_{\mathcal{X}})$ and $(\mY, \rho_{\mathcal{Y}})$, respectively,  denote premetric spaces of negative type. 
Then, the {\it generalized distance covariance} of two random elements $X\in\mX$ and $Y\in\mY$ with $\E |\rho_\mX(X,X')+\rho_\mY(Y,Y')| < \infty$ is defined as:
\begin{equation}
		\label{eq:gendcov2}
		\V^2_{\rho_\mX,\rho_\mY}(X,Y) = \E\Big[\rho_\mX(X,X') \left\{\rho_\mY(Y,Y')-\rho_\mY(Y,Y'')\\ -\rho_\mY(Y',Y'')+\rho_\mY(Y'',Y''')\right\}\Big], 
\end{equation}
which is clearly reminiscent of Equation~\eqref{eq:dcovalt}.


Consider now i.i.d.~joint samples $\bX = (X_1,\ldots,X_n)$ and $\bY = (Y_1,\ldots,Y_n)$ of $(X,Y)$, and define the distance matrices for each sample: $\bD^{\bX}:=\left( \rho_\mX(X_i,X_j)\right) _{n\times n}$  and $\bD^{\bY} := \left( \rho_\mY(Y_i,Y_j)\right)  _{n\times n}$. Then their doubly-centered versions $\tilde{\bD}^{\bX}$ and $\tilde{\bD}^{\bY}$ can be computed as follows:
\begin{equation} \label{eq:doublecenter}
\tilde{\bD}^{\bX} = (I-H) \bD^{\bX} (I-H) , \quad \quad  \tilde{\bD}^{\bY} = (I-H) \bD^{\bY} (I-H);
\end{equation}
where $H= \frac{1}{n} \boldsymbol{1} \boldsymbol{1}^t\in\mathbb R ^{n\times n}$ and $\boldsymbol{1}$ is an $n$-vector of ones. With this notation, a consistent empirical estimator of~\eqref{eq:gendcov2} can be written as,

\begin{equation}
\label{eq:dcovemp}
\widehat{\V}^2_{\rho_\mX,\rho_\mY}(\bX,\bY) = \frac{1}{n^2} \sum_{i,j=1}^{n} \tilde{\bD}^{\bX}_{i,j} \tilde{\bD}^{\bY}_{i,j}  = 
\frac{1}{n^2} 
\tr ({\bD}^{\bX} \tilde{\bD}^{\bY})=
\frac{1}{n^2} 
\tr (\tilde{\bD}^{\bX}{\bD}^{\bY}).
\end{equation}

\section{GENERALIZED DISTANCE COVARIANCE FOR TESTING THE ASSOCIATION OF A SINGLE SNP WITH A CONTINUOUS RESPONSE}\label{sec:testing}

\subsection{Tailoring premetrics to SNP data}

In this section, we investigate versions of generalized distance covariance $\V_{\rho_\mX,\rho_\mY}$ for testing independence between a SNP $X \in \mX := \{0,1,2\}$ and a quantitative response $Y \in \R$. As elucidated in Section \ref{gdc}, $\V_{\rho_\mX,\rho_\mY}$ is fully specified by choosing premetrics $\rho_\mX$ and $\rho_\mY$ on $\mX$ and $\mY$, respectively. While many distances on $\R$ appear sensible, we restrict ourselves to $\rho_\mY(y,y') = \frac{1}{2} |y-y'|^2$ since it leads to both tractable test statistics (Section \ref{sec:testing}) and illustrative interpretations (Section \ref{GT}). 

For defining meaningful distances on the support space of the SNPs, we note that $0$ and $2$ correspond to homozygous states, while $1$ denotes the heterozygous state. The definition which homozygous state is $0$ and which is $2$ is typically given by the convention of using $0$ for the homozygous state $A_1 A_1$ of the more frequent allele $A_1$. This convention seems arbitrary and leads to different codings of the homozygous states in different datasets. We argue that any reasonable testing procedure should be invariant to this convention. 
Consequently we only consider premetrics for which the distances between the heterozygous state and each of the homozygous states are equal. In the following, we hence set $d_b(0,1)=d_b(1,2) = 1$; 
note that the conclusions of the test would be the same under any scale transformations.

The resulting family of distances is characterized by the nonnegative real number $b:=d(0,2)$. For a premetric to define a distance covariance in the sense of Section \ref{gdc}, it must be of negative type and for this in turn, its square root must satisfy the triangle inequality  (cf. Section \ref{gdc}). This holds if and only if $\sqrt{d(0,2)} \leq \sqrt{d(0,1)} + \sqrt{d(1,2)} = 2$, which is equivalent to $b \leq  4$. 
For $b=0$, $d$ does not define a semimetric in the sense of \cite{Sejdinovic}, since two distinct points have distance $0$. However, as demonstrated by \cite{DJ}, the theory by \cite{Sejdinovic} readily extends to premetrics, assimilating points that are separated with distance zero (i.e., dropping the identity of indiscernibles).

We will hence study the family of premetrics $\{d_b\}_{b\in[0,4]}$, where $d_b:\spx\times\spx\longrightarrow\R$ is such that $d_b(0,1) = d_b(1,2) =1$ and $d_b(0,2)=b \in [0,4]$. Important special cases are:
\begin{itemize}
	\item The discrete metric $d_1(0,1) = d_1(1,2) = d_1(0,2) = 1$, as studied in \citet{F:epistasis} and \citet{F:categorical}.
	\item The absolute distance $d_2(x,x') = |x-x'|$, connected to standard distance covariance on the ordered set $\{0,1,2\} \subset \mathbb{R}$.
	\item The squared distance $d_4(x,x') = (x-x')^2$, linked to linear regression on the ordered set $\{0,1,2\} \subset \R$ (cf. \cite{DJ}).
\end{itemize}

\begin{figure}
    \centering
    \includegraphics[width=\textwidth]{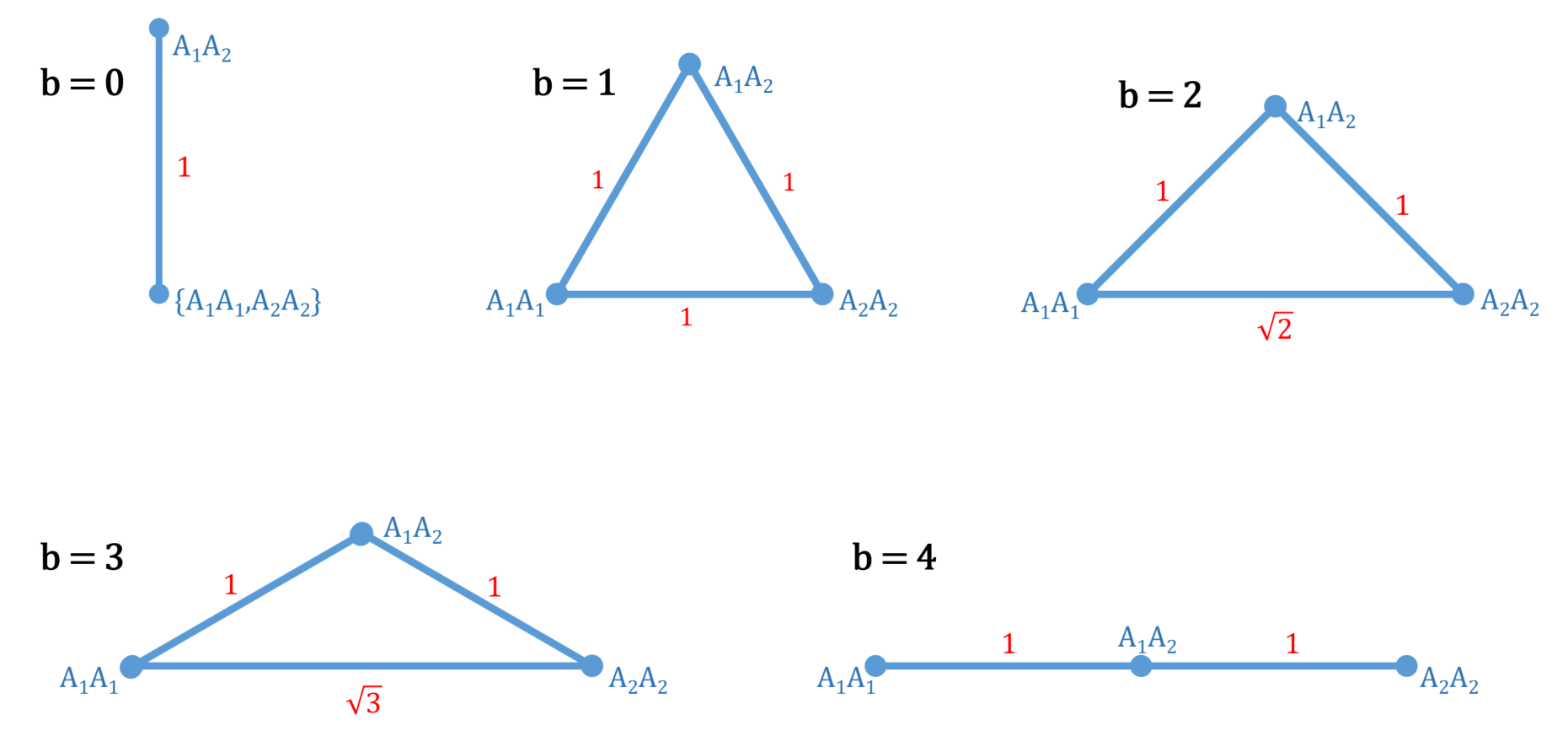}
    \caption{$\sqrt{d_b(\cdot,\cdot)}$ (in red) for several choices of $b$ between the first homozygous state $A_1 A_1 \,  \widehat{=} \, 0$, the heterozygous state $A_1 A_2 \, \widehat{=} \,1$ and the second homozygous state $A_2 A_2 \, \widehat{=} \, 2$. }
    \label{fig:metric}
\end{figure}

We also note that any premetric $d_b$ with $b \in (1,4)$ is related to the $\alpha-$distance covariance of \citep{SRB} for $\alpha = \log_2 b$. Figure \ref{fig:metric} illustrates some of the metrics by plotting $\sqrt{d_b(\cdot,\cdot)}$ (since the triangle inequality only needs to hold for the squareroot and $d_4$ corresponds to standard linear regression treating $X \in \{0,1,2\}$ as continuous, it appears more sensible to plot $\sqrt{d_b(\cdot,\cdot)}$ than $d_b(\cdot,\cdot)$).

Consider now the classical genotype-phenotype association models introduced in Section \ref{Models:SNP}. If the association model is known beforehand, it appears sensible to tailor the distance $d_b$ on the genotype level to the model under consideration. In particular, it is easy to see (cf. Figure \ref{fig:metric}) that the distance $d_0$ is tailored to a purely heterozygous model, where $\mu_0 = \mu_2$; one similarly obtains  that $d_4$ is a sensible choice for the additive model with heterozygous effect $h= \frac{\mu_1 - \mu_0}{\mu_2 - \mu_0} = \frac{1}{2}$. In an analogous fashion, one can define distances tailored to partially dominant models and overdominant models with arbitrary heterozygous effect $h$. For example using the distance $\rho$ with $\rho(0,1) = c\, h^2$, $\rho(1,2) = c \, (1-h)^2$, $\rho(0,2) = c$ for any constant $c > 0$ leads to a version of distance covariance tailored to partially dominant models with heterozygous effect $h$. 

However, the exact genotype-phenotype association model is typically unknown in practice, and we will see in the following that it is precisely for this situation that the GDC based on $d_b$ ($b \in (0,4)$) shows its strengths. In particular, we will see in Section \ref{GT} that each $d_b$ corresponds to the locally most powerful tests in a specific situation with uncertain association model; this uncertainty can for example be expressed via uncertainty of the heterozygous effect parameter $h$, where the distribution of $h$ is determined via $b$.

For the rest of the article, we use the simplified notation
    $$
     {\V}_b := {\V}_{d_b,\rho_\mY}, \quad  \widehat{\V}_b := \widehat{\V}_{d_b,\rho_\mY}, 
    $$
where $\rho_\mY(y,y') = \frac{1}{2} |y-y'|^2$. 
Unlike classical distance covariance, ${\V}_b$ does not characterize independence because $\rho_\mY$ is not of strong negative type. However, ${\V}_b$ can detect all associations defined via the classical phenotype-genotype association models introduced in Section \ref{Models:SNP}. For this purpose, we again consider
$$
    \mu_j = \E[Y|X=j]
$$
for $j\in\spx = \{0,1,2\}.$ Moreover, we define
$$
    p_j = \P(X=j).
$$
Then if $b \in (0,4)$ and the first moment of $Y$ exists, we have that the distance covariance between $X$ and $Y$ vanishes if and only if the mean of $Y$ is homogeneous among the values of $X$, i.e. if and only if $\mu_0=\mu_1=\mu_2$.

\begin{theorem} \label{th:dcovzero}
    Let $(X,Y)$ be jointly distributed random variables in $ \{0,1,2\} \times \R$ with $\E[Y] < \infty$. If $\mu_0 = \mu_1 = \mu_2$, then
    $$
        \V_b^2(X,Y) = 0.
    $$
 Moreover, if $b \in (0,4)$ and $p_j >0$ for $j \in \{0,1,2\}$, then $\mu_i \neq \mu_j$ for some $i\neq j$ implies that
     $$
        \V_b^2(X,Y) > 0.
    $$
\end{theorem}

The second part of Theorem \ref{th:dcovzero} does not hold true for the ``boundary cases'' of $\V_b$ with  $b \in \{0,4\}$; these are exactly the cases where $\widehat{\V}_b$ is tailored to one single genetic model (the purely heterozygous model for $b=0$ and the additive model for $b=4$.

\begin{proposition}
Let $b$ be either $0$ or $4$ and let $X$ be a random variable on $\{0,1,2\}$ with $p_j >0$ for $j \in \{0,1,2\}$. Then we can define a random variable $Y$ on the same underlying probability space such that $\mu_i \neq \mu_j$ for some $i\neq j$, but $\V^2_b(X,Y) = 0$.
\end{proposition}

Theorem \ref{th:dcovzero} implies that, for $b \in (0,4)$, the empirical version $\widehat{\V}_b$ may be used to establish consistent tests for 
the null hypothesis
    $$
        H_0: \mu_0 = \mu_1 = \mu_2.
    $$

In the following, we will introduce tests based on the asymptotic and the finite sample distribution of $\widehat{\V}_b$.

\subsection{Asymptotic and finite sample distribution}


The asymptotic distribution of  distance covariance is known to follow a infinite weighted sum of chi-square distributed random variables \citep{SRB,ETR} . While some approaches for approximating this distribution via moment-matching \citep{Berschneider, Huang} or approximation of the asymptotic distribution \citep{Fischer:kernel} have been proposed, the predominant procedure for testing is still to use resampling methods. However, resampling is hardly feasible in the GWAS setting, where a large number of tests have to be performed.

In the following, we derive a closed-form expression for our version of generalized distance covariance $\widehat{\V}_b$, which enables fast testing in GWAS.

\begin{theorem}\label{testasy}
    Let $\bX=(X_1,\ldots,X_n)$ and $\bY = (Y_1,\ldots,Y_n)$ denote i.i.d. samples of jointly distributed random variables $(X,Y) \in \{0,1,2\} \times \R$ with $Var(Y) = \sigma_Y^2 <\infty$. If $X$ and $Y$ are independent, then, for $n \to \infty$,
    $$
        n \, \widehat{\V}_b^2 \stackrel{\mathcal{D}}{\longrightarrow}  
    \sigma_Y^2 (\lambda_1 Q_1^2 + \lambda_2 Q_2^2),
        $$
where $Q_1^2$ and $Q_2^2$ are chisquare distributed with one degree of freedom and $\lambda_1$ and $\lambda_2$ are the eigenvalues of the matrix 
    $$
        K = \begin{pmatrix} \frac{b}{2} (p_0 + p_2 - (p_2 -p_0)^2) & \sqrt{\frac{b \, (4-b)}{4}} p_1 (p_0-p_2)\\
        \sqrt{\frac{b \, (4-b)}{4}} p_1 (p_0-p_2) & \frac{4-b}{2} (p_1-p_1^2)\end{pmatrix}.
    $$
  \end{theorem}

Using the asymptotic distribution for testing is typically more problematic in GWAS than for standard settings, since the convergence is  slower in the tails of the distributions and we are often interested in approximating very small p-values.

Assuming that the phenotype for each of the three genetic states is normally distributed with homogeneous variance, we can derive the finite-sample distribution of $\widehat{\V}_b^2$.
  
  \begin{theorem} \label{testfinite}
    For $n \in \mathbb{N}$, let $\bX=(X_1,\ldots,X_n) \in \{0,1,2\}^n$ denote a fixed sample and let $\bY = (Y_1,\ldots,Y_n)$ be defined by
        $$
         Y_i = \mu_j \, 1_{\{X_i = j\}} + \varepsilon_i,
        $$
where $\boldsymbol{\mu} = (\mu_0, \mu_1, \mu_2)^t \in \R^3$  and $(\varepsilon_1,\ldots,\varepsilon_n)$ is i.i.d. with $\varepsilon_i \sim \mathcal{N}(0,\sigma^2_Y)$. If $\mu_0 = \mu_1 = \mu_2$, then,
\begin{equation}\label{eq:testfinite1}
        \mathbb{P} \left( \frac{n \, \widehat{\V}_b^2}{\widehat{\sigma}_Y^2} \geq k \right) = P(T_n \geq 0),
\end{equation}
where $\widehat{\sigma}_Y^2 = \frac{1}{n} \sum_{j=1}^n (Y_j -\tfrac{1}{n} \sum_{i=1}^n Y_i)^2 $,
$T_n$ is defined by
\begin{equation} \label{eq:testfinite2}
        T_n = \Big(\widehat{\lambda}_1- \frac{k}{n} \Big) \, Q_1^2 + \Big(\widehat{\lambda}_2- \frac{k}{n} \Big) \, Q_2^2 - \frac{k}{n} Q_3^2 - \cdots - \frac{k}{n} Q_{n-1}^2
\end{equation}
and $Q_1^2,\ldots,Q_{n-1}^2$ are i.i.d. chisquare distributed with one degree of freedom; $\widehat{\lambda}_1$ and $\widehat{\lambda}_2$ are the eigenvalues of the matrix 
    $$
        K = \begin{pmatrix} \frac{b}{2} (\widehat{p}_0 + \widehat{p}_2 - (\widehat{p}_0 -\widehat{p}_2)^2) & \sqrt{\frac{b \, (4-b)}{4}} (-\widehat{p}_1 (\widehat{p}_0-\widehat{p}_2) \\
        \sqrt{\frac{b \, (4-b)}{4}} (-\widehat{p}_1 (\widehat{p}_0-\widehat{p}_2) & \frac{4-b}{2} (\widehat{p}_1-\widehat{p}_1^2)\end{pmatrix},
    $$
where 
$$
    \widehat{p}_i = \frac{1}{n} \sum_{j=1}^n 1_{\{X_j = i \}}, \quad i \in \{0,1,2\}.
$$
  \end{theorem}



\subsection{Computing \textit{p}-values} \label{sec:pvalues}

In GWAS, often very small test sizes are used, the standard being the so-called genome-wide significance threshold of $\alpha = 5 \times 10^{-8}$ \citep{Risch}. In this setting, using the asymptotic results leads to notable inflation of the type I error rate even for moderately large sample sizes. For this reason, we recommend to use the finite-sample distribution in Theorem \ref{testfinite}, except for very large sample sizes, say $n > 30,000$.

For calculating p-values, we first observe that by Theorem \ref{testfinite}, for $\widehat{\lambda}_2- \frac{k}{n} > 0$,
\begin{align*}
     \mathbb{P} \left( \frac{n \, \widehat{\V}_b^2}{\widehat{\sigma}_Y^2} \geq k \right)&= \mathbb{P} \left( \frac{\big(\widehat{\lambda}_1- \frac{k}{n} \big) \, Q_1^2 + \big(\widehat{\lambda}_2- \frac{k}{n} \big) \, Q_2^2}{\frac{1}{n-3} (Q_3^2 - \cdots - Q_{n-1}^2)} \geq \frac{k (n-3)}{n} \right) \\ &=1 - G_{F(2 (\widehat{\lambda}_1- \frac{k}{n}), 2 (\widehat{\lambda}_2- \frac{k}{n}); n-3)} \left(\frac{k (n-3)}{n} \right),
\end{align*}

where $G_{F(\alpha_1,\alpha_2; \nu)}$ is the cumulative distribution function of a generalized F-distribution in the terminology of \cite{Ramirez}. A closed-form expression for $G_{F(\alpha_1,\alpha_2; \nu)}$  can be derived from the general result in \cite{Dunkl},
\begin{equation} \label{eq:appell}
    G_{F(\alpha_1,\alpha_2; \nu)}(x) =  \,\Big(\frac{\nu \alpha_2}{2 x+ \nu \alpha_2} \Big)^{\nu/2+1} \, \frac{x}{ \sqrt{\alpha_1 \alpha_2}} F_1 \left( \frac{\nu}{2} +1, \frac{1}{2}, 1; 2; \frac{(1-\frac{\alpha_2}{\alpha_1})x}{(x+ \frac{\nu \alpha_2}{2})} , \frac{x}{(x+ \frac{\nu \alpha_2}{2})} \right),
\end{equation}
where $F_1$ is the Appell $F_1$ hypergeometric series \citep{Appell}.
 The test described in Theorem \ref{testfinite} can be regarded as a generalization of the classical F-test in linear regression. In particular, for $b=0$ and $b=4$, it follows that $\widehat{\lambda}_2 = 0$ and we obtain exactly the F-statistic from a simple linear regression model with predictors $1_{\{X=1\}}$ (corresponding to a purely heterozygous model) and $X$ (corresponding to an additive model), respectively.

For calculating the p-value, we can either numerically evaluate the closed form expression using efficient algorithms for the Appell $F_1$ hypergeometric series  or use one of the many algorithms suited for weighted sums of chi-square distributed variables using Equations \eqref{eq:testfinite1} and \eqref{eq:testfinite2}  \citep{Duchesne}. From our experience the former option is both computationally more efficient and more precise, so we use it as default option and for all our calculations in this paper. While the main part of the code is written in R, we use the \texttt{reticulate} to call the python package \texttt{mpmath} for a precise and computationally efficient calculation of the Appell $F_1$ hypergeometric. To further speed up the calculation, we now derive upper and lower bounds for the p-values derived from the distribution in Theorem \ref{testfinite} .

\begin{proposition}
Let $G_{\chi^2(w_1,w_2)}$ denote the cumulative distribution function of the random variable $w_1 Q_1^2 + w_2 Q_2^2$, where $Q_1^2$ and $Q_2^2$ are i.i.d. chisquare distributed with one degree of freedom. Further, let $G_{F(d_1,d_2)}$  denote the cumulative distribution function of the classical $F$-distribution with $d_1$ and $d_2$ degrees of freedom.   Then
    $$
 p^*  \leq \mathbb{P} \left( \frac{n \, \widehat{\V}_b^2}{\widehat{\sigma}_Y^2} \geq k \right) \leq p^{**},
    $$    
where for $\widehat{\lambda}_2 - \frac{k}{n} >0$, 
    $$
  p^* = 1- \min\left( G_{\chi^2(\widehat{\lambda}_1- \frac{k}{n}, \widehat{\lambda}_2- \frac{k}{n} )} \left(\frac{k (n-3)}{n} \right) , G_{F(1,n-3)} \left(\frac{k (n-3)}{\widehat{\lambda}_1 n - k} \right), G_{F(2,n-3)} \Bigg(\frac{k (n-3)}{ \prod_{i=1}^2 (\widehat{\lambda}_i n - k)^{1/2} \, } \Bigg)\right)
    $$    
and 
    $$
p^{**} =  5 \, \left( 1- G_{F(1,n-2)} \left(\frac{k (n-2)}{\widehat{(\lambda}_1 + \widehat{\lambda}_2) n - 2 \, k } \right) \right).
    $$
For $\widehat{\lambda}_2 - \frac{k}{n} \leq 0$, 
    $$
  p^* = 1-  G_{F(1,n-2)} \left(\frac{k (n-2)}{\widehat{\lambda}_1 n - k} \right), \, \text{and} \quad   p^{**} = 1-  G_{F(1,n-3)} \left(\frac{k (n-3)}{\widehat{\lambda}_1 n - k} \right).
    $$  
\end{proposition}

When performing GWAS in practice, it is usually not interesting to calculate the precise number for p-values substantially greater than the genome-wise significant threshold (say greater than $M =  10^{-3}$). On the other hand, it may also be not sensible to precisely evaluate extremely small p-values (say smaller than $m = 10^{-32}$). 

For a fast algorithm, we first calculate the approximations $p^*$ and $p^{**}$ for all SNPs. This can be carried out extremely efficiently; e.g. with help of the \texttt{R} packages \texttt{stats} and \texttt{coga}. Precise evaluation of the p-values in Theorem \ref{testfinite} is then only performed for the SNPs satisfying $p^* < M$ and $p^{**} > m$. In Section  \ref{sec:simu}, the speed of this fast algorithm is compared to the naive algorithm, which evaluates the precise p-value for all SNPs. It is demonstrated that the use of the fast algorithms leads to substantial gains in computational efficiency.



\section{LOCALLY MOST POWERFUL PROPERTY AND INTERPRETATIONS}\label{GT}

In Section \ref{sec:testing}, we have derived a computationally efficient test that can detect all  alternatives that can be expressed by the classical genetic associations in \cite{GI} that we have introduced in Section \ref{Models:SNP}. In the following, we show that for each $b \in [0,4]$, $\widehat{\V}_b^2$ features a valuable interpretation as the locally most powerful test statistic in certain statistical models. This provides both a theoretical guarantee for the statistical efficiency of $\widehat{\V}_b^2$ and provides a better understanding, which choices of $b$ are suitable from a biological viewpoint.

The classical score test \citep{cox1979theoretical} for a model with likelihood $\ell^{*}(\theta; \bZ)$ where $\bZ \in \R^n$ is an observation and $\theta \in \Theta \subset \R$ is a univariate parameter, is a one-sided test $H_0^* : \theta = \theta_0$ against
$H_1^* : \theta > \theta_0$ that rejects $H_0^*$ if
	$$
	 S^* = \frac{d\log \ell^*(\theta_0; \bZ)}{d \theta} \geq c
	 $$
for some critical value $c$. The score test is also known as the {\it locally most powerful test} since it satisfies the following optimality property

\begin{lemma}[\cite{goeman2006testing}, Lemma 2] \label{lem:goeman}
For $\theta \in \Theta$, denote by $Z_\theta \in \R^n$ a random variable distributed corresponding to $\ell^{*}(\theta; \bZ)$ and denote its probability measure by $P_{\theta}$. Suppose that the derivative $\frac{d \ell^*(\theta; \bZ)}{d \theta}$ exists for all $Z \in \R^n$ and is bounded in a neighbourhood of $\theta_0$. Then, for any test of $H_0^*$ with critical region $A$ and power function $w(\theta) = P_{\theta}(A)$, the derivative $\frac{d w (\theta_0)}{d \theta}$ exists. Also, denote the power function of the score test statistic by $w^* (\theta) = P_{\theta} (S^* \geq c)$ for some $c \geq 0$. Then $$
    w(\theta_0) \leq w^*(\theta_0)
$$
implies
$$
\frac{d}{d \theta} w(\theta_0) \leq  \frac{d}{d \theta} w^*(\theta_0).
$$
\end{lemma}
\noindent Since
    $$
        P_{\theta_0+h} (A) = w(\theta_0+h) 
        = w(\theta_0) + h \, \frac{d}{d \theta} w(\theta_0) + o(h),
    $$
Lemma \ref{lem:goeman} implies that no test of the same size can be more powerful for infinitesimally small deviations from $\theta_0$. This implies that the score test is the most powerful test for detecting local alternatives corresponding to infinitesimally small deviations from $\theta_0$ or short {\it locally most powerful test}.

\cite{DJ} have shown that, if the squared Euclidean distance is applied on the response, the generalized distance covariance arises from the score test statistic in certain Gaussian regression models.
This implies that $\widehat{\V}^2_b$ has an interpretation as locally most powerful test statistic, which we state in Theorem \ref{th:lomopo} and Remark \ref{rem:lomopo}.

For a better understanding of the results of the following sections, it is helpful to display the squared GDC $\widehat{\V}_b^2$ as a version of the Hilbert--Schmidt Independence Criterion (HSIC) introduced by \cite{gretton2008kernel}. Following \cite{Sejdinovic}, the squared GDC $\V^2_{\rho_\mX,\rho_\mY}$ can be written as the Hilbert--Schmidt Independence Criterion $\mbox{HSIC}_{k_\mX, k_\mY}$, where (with an arbitrary point $x_0 \in \mX$),
 \begin{equation}
	\label{eq:smindkernel}
	k_\mX (x_1,x_2) = \rho_\mX(x_1,x_0) + \rho_\mX(x_2,x_0) -  \rho_\mX(x_1,x_2),
	\end{equation}   
 $k_\mX$ is then called the kernel {\it induced} by $\rho_\mX$.
To define the (empirical)  HSIC for samples $\bX$ and $\bY$ of jointly distributed random variables $(X,Y)$, we construct kernel matrices $\bK^{\bX}=\left( k_\mX(X_i,X_j)\right) _{n\times n}$ and $\bK^{\bY}$ and double-center them analogous to Eq. \eqref{eq:doublecenter} (i.e. $\tilde{\bK}^{\bX} = (I-H) \bK^{\bX} (I-H)$); then HSIC is defined as,
\begin{equation}
	\label{eq:HSICemp}
	\widehat{\textrm{HSIC}}_{k_\mX,k_\mY}(\bX,\bY) = \frac{1}{n^2} \sum_{i,j=1}^{n} \tilde{\bK}^{\bX}_{i,j} \tilde{\bK}^{\bY}_{i,j} = \frac{1}{n^2} \tr ({\bK}^{\bX} \tilde{\bK}^{\bY})= \frac{1}{n^2} \tr (\tilde{\bK}^{\bX} {\bK}^{\bY}),
\end{equation}
a population version of HSIC can be defined analogously to \eqref{eq:gendcov2} (or as the a.s. - limit of \eqref{eq:HSICemp}). 

The following proposition provides kernels induced by the family of distances $d_b$.

\begin{proposition} \label{prop:kernel} 
The distance $d_b$ induces the kernel $k_b$ with
    $$
k_b(0,0) = k_b(2,2) = 1; \quad k_b(1,1) = k_b(0,1) = k_b(1,2) = 0; \quad k_b(0,2) = 2-b.
    $$
\end{proposition}

Under certain assumptions, a kernel $k: \mZ \times \mZ \to \R$ can be decomposed into {\it features}, that is, there exists a so-called feature map $\boldsymbol{\Phi}: \mZ \to \R^p$ such that $$k(z,z') = \langle \boldsymbol{\Phi}(z), \boldsymbol{\Phi}(z') \rangle,$$
where $\langle \cdot, \cdot \rangle$ denotes the standard inner product in $\R^p$. Throughout this work, we will say short ``{\it feature map of a (pre-)metric $\rho$}'' for a feature map of a kernel $k$ which is induced by a (pre-)metric $\rho$.

\begin{proposition} \label{prop:fm}
A  feature map $\boldsymbol{\Phi}=(\phi_1,\phi_2)$ of $d_b$ is given by
$$
\phi_1(x) = \sqrt{\frac{b}{2}} (-1_{\{x=0\}} + 1_{\{x=2\}}) , \quad \phi_2(x) = \sqrt{\frac{4-b}{2}} 1_{\{x=1\}}
$$
or in vector notation (that we will use throughout the manuscript),
   $$
        \phi_1 = \sqrt{\frac{b}{2}} \begin{pmatrix} -1 \\ 0 \\1 \end{pmatrix}, \, \phi_2 = \sqrt{\frac{4-b}{2}} \begin{pmatrix} 0 \\ 1 \\ 0 \end{pmatrix}.
    $$
\end{proposition}

\medskip

Using its HSIC representation (cf. Equations \eqref{eq:smindkernel} and \eqref{eq:HSICemp}), $\widehat{\V}_b^2$ can alternatively be written as
\begin{equation} \label{eq:HSICstat}
    \widehat{\V}^2_b(\bX, \bY) = \frac{1}{2 \, n^2} \sum_{i,j=1}^n k_b(X_i,X_j) (Y_i - \widehat{\mu}_Y) \, (Y_j - \widehat{\mu}_Y)
\end{equation}
with 
$ \widehat{\mu}_Y = \frac{1}{n} \sum_{j=1}^n Y_j$.

Theorem \ref{th:lomopo} provides an interpretation of $\widehat{\V}^2_b$ as the locally most powerful test statistic in a Gaussian regression model.

\begin{theorem} \label{th:lomopo}
	Let $(\phi_1,\ldots,\phi_r)$ be a feature map induced by the distance $d_b$ with corresponding kernel $k_b$, as e.g. provided by Proposition \ref{prop:fm}.
	Consider the model
		\begin{equation} \label{eq:lomopo}
		 Y_i = \sum_{j=1}^r \beta_j \phi_j (X_i) + \mu_Y + \varepsilon,
		\end{equation}
where $\mu_Y$ is known, $\varepsilon \sim \mathcal{N}(0,\sigma^2)$ and, for $j \in \{1,\ldots,r\}$, $\beta_j = \tau B_j$ with $\tau \in \R$ and $B_1,\ldots,B_r$ are mutually uncorrelated random variables with $\E[B_j] = 0$ and $\E[B_j^2]=1$. Then the locally most powerful test statistic for testing 
    $$
        H_0 : \tau^2 = 0 \text{ against } H_1 : \tau^2 > 0
    $$
is  given by
 \begin{equation} \label{eq:lomopo}
 	\widehat{\mathcal{U}}_b^2 =  \frac{1}{n^2}	\sum_{i,j=1}^n k_b(X_i,X_j) (Y_i- \mu_Y) (Y_j-\mu_Y).
\end{equation}
\end{theorem}

\begin{remark} \label{rem:lomopo}   
The population mean $\mu_Y$ is typically unknown in practice.
	By plugging in the sample mean $\widehat{\mu}_Y$ for $\mu_Y$ in \eqref{eq:lomopo}, we see that a pivot statistic for $\widehat{\mathcal{U}}_b^2$ is given by the squared generalized distance covariance $\widehat{\V}_b^2$ in \eqref{eq:HSICstat}.
\end{remark}

In GWAS, it is usually conjectured that the effect of a single SNPs on a quantitative trait is small, hence the assumption of a small $\tau$ appears sensible. Consequently the locally most powerful property is particularly desirable for this setting. 
Theorem \ref{th:lomopo}  does neither specify the exact distribution of  $(B_1,\ldots,B_r)$ nor the feature map $(\phi_1,\ldots,\phi_r)$. In the following section, we demonstrate that different choices of  $(B_1,\ldots,B_r)$ and $(\phi_1,\ldots,\phi_r)$ lead to different interesting interpretations of Theorem \ref{th:lomopo}, providing insights into the nature of  $\widehat{\V}_b^2$; in particular it will be useful to consider feature maps that are different from the one provided in Proposition \ref{prop:fm} and have more than two features.

\subsection{Interpretation as locally most powerful tests for situations in which dominant-recessive, additive or purely heterozygous models are expected for certain fractions of SNPs}\label{GT:mixtures}

For a first interpretation, we consider that the random vector $(B_1,\ldots,B_r)$ in Theorem \ref{th:lomopo} satisfies $\P(B_i \neq 0, B_j \neq 0) =0$ for $i \neq j$. This implies that only one of the coefficients $\beta_1, \ldots, \beta_r$ is nonzero and hence only one of the features in Equation \eqref{eq:lomopo} is involved for each realization of the model.

\begin{corollary} \label{cor:mix}
	Let $(\phi_1,\ldots,\phi_r)$ be a feature map induced by the distance $d_b$ and, for $j \in \{1,\ldots,r\}$, let $c_j >0$; further denote $\psi_j(\cdot) = \phi_j (\cdot) /c_j$.
	Let $U$ be a discrete random variable with $P(U = j) = \frac{c_j^2}{\sum_{k=1}^n c_k^2}$ and consider the model
		$$
		 y_i = \tau \,A \sum_{j=1}^r 1_{\{U=j\}} \psi_j (x_i) + \mu_Y + \varepsilon,
		$$
 where $\tau \in \R$, $\mu_Y$ is known, $\varepsilon \sim \mathcal{N}(0,\sigma^2)$ and $A$ is a random variable, independent of $U$ with $\E[A]=0$ and $0 < \E[A^2] < \infty$ (e.g. $P(A=1)=P(A=-1)=\tfrac12$). Then the locally most powerful test for testing 
    $
        H_0 : \tau^2 = 0 \text{ against } H_1 : \tau^2 > 0
    $ is given by \eqref{eq:lomopo}.
\end{corollary}

For facilitating interpretation, the factors $c_j$ should be chosen in a way such that the standardized features $\psi_j$ are on a comparable scale. The variable $A$ balances positive and negative effects of a feature (guaranteeing $E[B] = 0$ in Theorem \ref{th:lomopo}).

Corollary \ref{cor:mix} states that $\widehat{\V}_b^2$ is nearly (cf. Remark \ref{rem:lomopo}) the locally most powerful test statistic in a model, where each of $r$ different association patterns (specified by the $r$ standardized features of the feature maps) between a SNP $X$ and a quantitative response $Y$  is present with a certain probability. We note that this is different from a mixture model, in the sense that the random parameters $U$ and $A$  do not depend on $i$, but are only drawn once and hence the same model is true for all samples $i$. 

Considering that we would typically apply the same test for each of a large number of SNPs, the corresponding test is in a certain sense optimal for situations in which the association patterns expressed by $\psi_j$ shows up for a fraction of $\frac{c_j^2}{\sum_{i=1}^n c_i^2}$ of the SNPs. 


We now introduce new feature maps leading to a particularly helpful interpretation of Corollary \ref{cor:mix} : 
For $b \in [2,4]$, we easily see that that a feature map of $d_b$ is given by,
 $$
        \phi_1 = \sqrt{{4-b}} \begin{pmatrix} 0 \\  0 \\ 1\end{pmatrix}, \, \quad \phi_2 = \sqrt{{4-b}} \begin{pmatrix} 0 \\  1 \\ 1\end{pmatrix}, \, \quad \phi_3 = 2\, \sqrt{b-2} \begin{pmatrix} 0 \\ \tfrac12 \\ 1 \end{pmatrix} 
    $$
Applying Corollary \ref{cor:mix} with $c_1 = c_2 = \sqrt{{4-b}}, \, c_3 = 2\,\sqrt{b-2}$ yields that for $b \in [2,4]$, $\V_b^2$ is optimal for a setting where the absolute difference between the two homozygous states is $|\tau|$ (with small $\tau$) and the heterozygous state takes the value of each of the homozygous states with probability $\frac{4-b}{2 \,b}$ and the average of the two values with probability $\frac{4\,(b-2)}{2b}$. This corresponds to the situation, where a dominant and recessive model hold for a fraction of $\frac{4-b}{2 \,b}$ of the SNPs each, and an additive model holds for a fraction $\frac{4\,(b-2)}{2b}$ of the SNPs.

In particular, $\V_2$ is optimal if all SNPs associated with $Y$ follow a dominant-recessive model and each of the homozygous states is dominant for one half of the SNPs, cf. Figure \ref{fig:lomopo}. $\V_3$ on the other hand is optimal for a situation, where a dominant-recessive model is present with probability $\frac{1}{3}$ (for which each of the homozygous state is dominant with the same probability) and an additive model is present with probability $\frac{2}{3}$. As pointed out before, the extreme case $\V_4$ corresponds to the locally most powerful test in a purely additive model and is equivalent to the test statistic obtained from linear regression with the SNP $X \in \{0,1,2\}$ as single predictor, cf. Figure \ref{fig:lomopo}.

Similarly, for $b \in [0,2]$, we can derive the feature map
    $$
        \phi_1 = \sqrt{{b}} \begin{pmatrix} 0 \\  0 \\ 1\end{pmatrix}, \, \quad \phi_2 = \sqrt{{b}} \begin{pmatrix} 0 \\  1 \\ 1\end{pmatrix}, \, \quad \phi_3 = \sqrt{2-b} \begin{pmatrix} 0 \\ 1 \\ 0\end{pmatrix}.
    $$
Applying Corollary \ref{cor:mix} with $c_1 = c_2 = \sqrt{b}, \, c_3 = \sqrt{2-b}$ yields, that, for $b \in [0,2]$, $\V_b^2$ is optimal for situations where a dominant-recessive model is present with probability $\frac{2 \,b}{2+b}$ (in which each of the homozygous states is dominant with equal probability) and a purely heterozygous model is present with probability $\frac{2-b}{2+b}$. 
For, $\V_2$, $c_3$ is zero and we obtain the same interpretation as above. $\V_1$ is optimal for a situation in which two means are equal and for each $j \in \{0,1,2\}$, $\mu_j$ differs from the other two means for $\tfrac13$ of the associated SNPs. This model is agnostic in the sense that it does not make any difference between the states $0, 1, 2$, which is also clear from $d_1(0,1)= d_1(0,2) = d_1(1,2) = 1$, cf. Figure \ref{fig:metric}. For $b=0$, we obtain $c_1 = c_2 =  0$; hence $\widehat{\V}_0$ is optimal  for a purely heterozygous model - as we have pointed out before, the corresponding test statistic is equivalent to that obtained from a linear regression with predictor $Z_i = 1_{\{X_i=1\}}$.

\subsection{Interpretation as the locally most powerful test for partially dominant and overdominant models with random parameters}\label{GT:codom:overdom}

While it is common that the response values for the heterozygous state lie between the values of the two homozygous states, it seems rather unlikely that we encounter an exact additive model. Instead, the response values of the heterozygous state will typically lie closer to one of the homozygous states. A model which assumes that the response values for the heterozygous state lie somewhere between the response values  of the two homozygous states is referred to as a {\it partially dominant model}, as indicated in Section 2.1.

We will now show that, for $b \in (2,4]$, $\V^2_b$ can be interpreted as the locally most powerful test statistic in certain random partially dominant models. 
For this purpose, we first state the following alternative formulation of the locally most powerful property.

\begin{theorem} \label{th:lomopo2}
	Consider the distance $d_b$ and
	assume the model
	\begin{align*}
	Y_i = \begin{cases}  \mu_Y + \varepsilon, &\text{ if $x_i =0$}, \\
	\mu_Y + \beta_1 + \varepsilon,  &\text{ if $x_i =1$}\\
	 \mu_Y + \beta_1 + \beta_2 + \varepsilon	&\text{ if $x_i =2$,}
	\end{cases}
	\end{align*}
	where $\mu_Y$ is known, $\varepsilon \sim \mathcal{N}(0,\sigma^2)$ and $(\beta_1, \beta_2) = \tau B$ with $\tau \in \R$ and $B$ is a random variable with $E[B] = 0$ and 
	$$
	E[B B^t] =  c \, \begin{pmatrix} 1 & \frac{b}{2}-1 \\ \frac{b}{2}-1 & 1  \end{pmatrix},
	$$
	where $c$ is some constant.
 Then the locally most powerful test for testing 
    $
        H_0 : \tau^2 = 0 \text{ against } H_1 : \tau^2 > 0
    $ is given by \eqref{eq:lomopo}.
\end{theorem}

This yields the interpretation of $\widehat{\V}_b$ as locally most powerful test statistics in regression models with correlated regression parameters. For $b \in [0,2)$, the correlation between $\beta_1$ and $\beta_2$ is negative. In this case, we can choose $B$ in a way such that $\beta_1$ and $\beta_2$ always have opposing signs. For $b \in (2,4]$ on the other hand, the correlation between $\beta_1$ and $\beta_2$ is positive and hence we can choose $B$ in a way such that $\beta_1$ and $\beta_2$ always have the same sign.

Reminding us of the association models introduced in Section 2.1, we can interpret $\V_b$ with $b \in (0,2)$ as the locally most powerful test in an overdominant model with random heterozygous effect $H$. Analogously $\V_b$ with $b \in (2,4)$ can be interpreted as the locally most powerful test in a partially dominant model with random heterozygous effect $H$. 

By choosing $\beta_1, \beta_2$ as two-sided gamma distributions with same sign, we obtain the following corollary, providing a particularly helpful interpretation of $\V_b$ for $b \in (2,4)$.

\begin{corollary} \label{cor:beta}
	Consider the distance $d_b$ with $b \in (2,4)$ and
	assume the model
	\begin{align*}
	Y_i = \begin{cases}  \mu_Y + \varepsilon, &\text{ if $x_i =0$}, \\
	\mu_Y + \tau H A + \varepsilon,  &\text{ if $x_i =1$}\\
	 \mu_Y + \tau A + \varepsilon	&\text{ if $x_i =2$,}
	\end{cases}
	\end{align*}
	where $\mu_Y$ is known, $\tau \in \R$, $\varepsilon \sim \mathcal{N}(0,\sigma^2)$ and the heterozygous effect $H$ is beta-distributed with parameters $(\frac{b-2}{4-b}, \frac{b-2}{4-b})$. $A$ is a random variable, independent of $H$ with $\E[A]=0$ and $\E[A^2]=1$ (e.g. $P(A = 1) = P(A = - 1) = \frac{1}{2}$).
 Then the locally most powerful test for testing 
    $
        H_0 : \tau^2 = 0 \text{ against } H_1 : \tau^2 > 0
    $ is given by \eqref{eq:lomopo}.
\end{corollary}

\begin{figure}
    \centering
    \includegraphics[width=\textwidth]{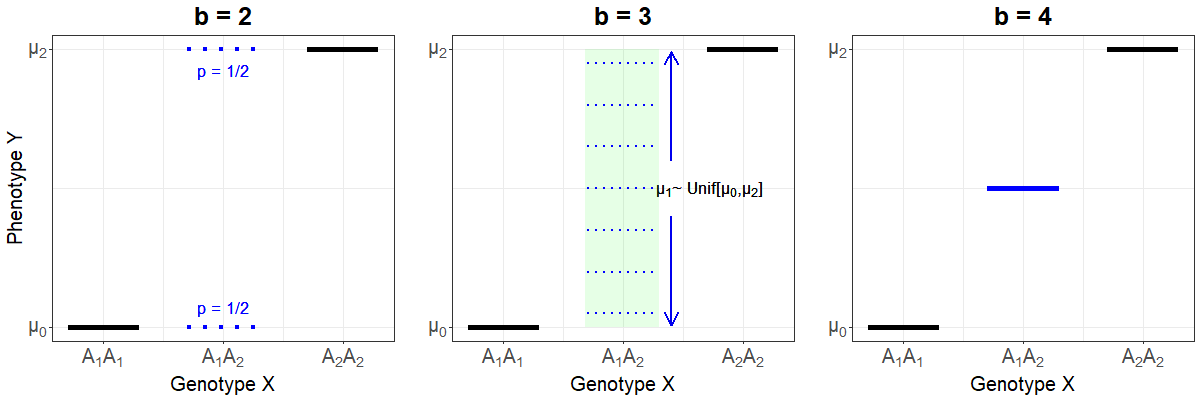}
    \caption{Illustration of the locally most powerful property for $\widehat{\V}_b$ with $b=2,3,4$. For $b=2$, the test is locally optimal for a dominant-recessive model, where each of the homozygous states is dominant with equal probability. For $b=3$, the test is locally optimal if the true mean of the response variable $Y$ is uniformly distributed on the interval bounded by the means of the homozygous states. For $b=4$, the test is locally optimal for an additive model.}
    \label{fig:lomopo}
\end{figure}

Corollary \ref{cor:beta} states that, for $b \in (2,4)$, $\V_b$ arises from the locally most powerful test in a partially dominant model for which the heterozygous effect parameter $H$ is beta-distributed with parameters $(\frac{b-2}{4-b}, \frac{b-2}{4-b})$. An important special case is $\V_3$, which is most powerful if the effect parameter $H$ is uniformly distributed on $[0,1]$ - i.e. $\V_3$ is optimal for a random Gaussian regression model where the mean $\mu_1$ of the response variable for the heterozygous state is uniformly distributed on the interval $[\mu_0,\mu_2]$ bounded by the means of the response variable for the homozygous states, see Figure \ref{fig:lomopo} for an illustration.

A similar result as in Corollary \ref{cor:beta} can be obtained for $b \in (2,4)$, see Appendix B in the Supplementary Material. We conclude this section with an overview of helpful interpretations for $\widehat{\V}_b$ for different parameter values $b$ in the range $[0,4]$ (Table \ref{tab:interpretation}). The most intuitive interpretations of the locally most powerful property for values $b=2,3,4$ are additionally illustrated in Figure \ref{fig:lomopo}.

\begin{table}[htbp] \label{tab:interpretation}
\caption{Genetic model against which $\widehat{\V}_b$ provides the locally most powerful test, for different values of $b\in[0,4].$ 
}
\begin{tabular}{l | l}
& Genetic model \\
\hline
$b=0$ & purely heterozygous model ($\mu_0 =\mu_2$)\\
$b \in (0,1)$ & overdominant model with large heterozygous effect ($h = \frac{G_1}{G_1 - G_2}$ with $G_i \sim \Gamma(\frac{2-b}{b},1)$) \\
$b=1$ & agnostic model, treating the states $\{0,1,2\}$ indifferently
 \\
$b \in (1,2)$ & overdominant model with small heterozygous effect ($h = \frac{G_1}{G_1 - G_2}$ with $G_i \sim \Gamma(\frac{2-b}{b},1)$) \\
$b = 2$ & dominant-recessive model with equal probability for dominance and recessiveness\\
$b \in (2,3)$ & partially dominant model where $h$ tends to be close to $0$ or $1$ ($h \sim \beta(\frac{b-2}{4-b}, \frac{b-2}{4-b})$)\\
$b =3$ & partially dominant model, heterozygous effect $h$ is uniformly distributed on $[0,1]$ \\
$b \in (3,4)$ & partially dominant model where $h$ tends to be close to $\tfrac12$ ($h \sim \beta(\frac{b-2}{4-b}, \frac{b-2}{4-b})$) \\
$b = 4$ & additive model, $h= \frac{1}{2}$
\end{tabular}
\end{table}

\section{ADJUSTING FOR NUISANCE COVARIATES}\label{Covariates}

In GWAS it is often necessary or beneficial to control for nuisance covariates. For an illustrative example, consider that we aim to test the association of a SNP $X$ with height $Y$ in adults including elderly individuals. Then it appears sensible to adjust for both sex and age, reducing variation in the response and leading to higher power. Moreover, the phenomenon known as \emph{population stratification} has been identified since the very beginning of the genomic era as a main cause of false positives in GWAS \citep{Cardon}. Since we know that the genetics of an individual is affected by the genetics of their parents, the ancestry of an individual may act as a confounder, as it influences both the genetics and the environment in which the person is born and grows \citep{Brandes}. Consequently it may be necessary to control for strata in the population, which may be done using information on ethnic groups or principal components of the full genomic information.

We now derive adjusted versions of $\V_b^2$ and $\widehat{\V}_b^2$ for testing in the presence of nuisance covariates. Different from other approaches \citep{Partial, Conditional}, we will adjust for the influence of the covariates in a linear fashion, which allows to retain both a tractable test statistic and a meaningful interpretation; nonlinear influences of the covariates can still be taken into account by transformations, using e.g. splines. 

For defining the linearly adjusted version of our GDC, let $Z = (1,Z_1,\ldots,Z_q)^t \in \R^{(q+1)}$ be a random vector with $\E [Z^2] < \infty$. Then define,
 $$
    \V^2_b(X\, , \, Y; \,Z) = \V^2_b(X, \, Y- Z^t \tilde{\gamma}),
 $$
where $\gamma$ is given by
$$
    \tilde{\gamma} = \argmin_{\gamma \in \R^{q+1}} (Y- Z^t \gamma)^2.
$$
Assuming that $\E[Y^2] < \infty$, we obtain the classical representation,
    $$
        \tilde{\gamma} = \E[Z Z^t]^{-1} E[Z Y]  .
    $$

The following corollary is an immediate consequence of Theorem \ref{th:dcovzero}.

\begin{corollary} \label{th:nuiscovpop}
If $\E[Y-\tilde{\gamma}^t Z| X=j] = 0$ for all $j \in \{0,1,2\}$, then 
    $$
        \V^2_b(X\, , \, Y; \,Z) = 0.
    $$
On the other hand, if $\E[Y-\tilde{\gamma}^t Z| X=j] \neq 0$ for some $j \in \{0,1,2\}$ and $p_j >0$ for all $j \in \{0,1,2\}$, then, if $b \in (0,4)$,
 $$
        \V^2_b(X\, , \, Y; \,Z) > 0.
    $$
\end{corollary}

\medskip

\noindent In particular, assuming $b \in (0,4)$, Corollary \ref{th:nuiscovpop} yields, that  in the setting of a linear regression,
    $$
Y = \widetilde{\gamma}^t Z  +\mu_j 1_{\{X=j\}} + \varepsilon ,
    $$
 $\V^2_b(X\, , \, Y; \,Z)$ ) equals $0$ if and only if $\mu_0 = \mu_1 = \mu_2$.

Given i.i.d. jointly distributed samples $\bX$, $\bY$, $\bZ$, we further define an empirical version,
$$
\widehat{\V}^2_b(\bX\, , \, \bY; \,\bZ) = \widehat{\V}^2_b(\bX, \, \bY-\bZ \widehat{\gamma} ),
$$
where $\widehat{\gamma}$ is the OLS estimate,
$$    
    \widehat{\gamma} = (\bZ^t \bZ)^{-1} \bZ^t \bY.
$$

Hence the adjusted version  $
\widehat{\V}^2_b(\bX\, , \, \bY; \,\bZ)$ is defined as the regular GDC $\widehat{\V}^2_b$ between $\bX$ and the residuals of a linear regression of $\bY$ on $\bZ$.

We now state the asymptotic distribution of $
\widehat{\V}^2_b(\bX\, , \, \bY; \,\bZ)$. Different from the case without covariates, naive resampling methods are not valid here, because the samples $(X_i, Y_i - Z_i^t \widehat{\gamma})$ are nonexchangeable. Hence, the derivation of the test statistic distribution is of utmost importance even for the case where we only consider a small number of SNPs.


\begin{theorem} \label{th:covasy} Let $Z = (1,Z_1,\ldots,Z_q)^t \in \R^{q+1}$ and $X \in \{0,1,2\}$ be random variables with $\E[Z^2] < \infty$. Assume the model
    $$
        Y =  \gamma^t Z + \mu_j 1_{\{X=j\}} + \varepsilon,
    $$
where $\varepsilon \in \R$ is independent of $(X,Z)$ with $\E[\varepsilon] = 0$, $\E[\varepsilon^2] = \sigma_\varepsilon^2 < \infty$ and $(\mu_0,\mu_1,\mu_2) \in \R^3$.  Further assume that $Z$ is non-singular. Consider now i.i.d. jointly distributed samples $\bX \in \{0,1,2\}^n$, $\bY \in \R^n $ and $\bZ \in \R^{n \times (q+1)}$ of $(X,Y,Z)$.

If $\mu_0 = \mu_1 = \mu_2$, then, for $n \to \infty$,
    $$
        n \, \widehat{\V}_b^2 (\bX, \, \bY; \, \bZ) \stackrel{\mathcal{D}}{\longrightarrow}  
        \sigma_\varepsilon^2 (\lambda_1 Q_1^2 + \lambda_2 Q_2^2),
        $$
where $Q_1^2$ and $Q_2^2$ are chisquare distributed with one degree of freedom and $\lambda_1$ and $\lambda_2$ are the eigenvalues of the matrix 
    $$
        K = E[\Phi(X) \Phi(X)^t] - E[\Phi(X) Z^t] \, (E[Z Z^t])^{-1}  E[\Phi(X) Z^t]^t   $$
and $\Phi(X) = (\phi_1(X),\ldots, \phi_r(X))^t$ with an arbitrary feature map $\phi_1(X),\ldots, \phi_r(X)$ of $d_b$.
\end{theorem}

Under Gaussianity, we can again derive the exact finite-sample distribution.

\begin{theorem} \label{th:covfinite} 
 For $n \in \mathbb{N}$, let $\bX=(X_1,\ldots,X_n) \in \{0,1,2\}^n$ denote a fixed sample and let $\bY = (Y_1,\ldots,Y_n)$ be defined by
        $$
         Y_i = \gamma^t Z_i + \mu_j \, 1_{\{X_i = j\}} + \varepsilon_i,
        $$
where $\boldsymbol{\mu} = (\mu_0, \mu_1, \mu_2)^t \in \R^3$, $Z_i \in \R^p$  and $(\varepsilon_1,\ldots,\varepsilon_n)$ is i.i.d. and independent of $(\bX,\bZ)$ with $\varepsilon_i \sim \mathcal{N}(0,\sigma^2_\varepsilon)$. If $\mu_0 = \mu_1 = \mu_2$, then,
    $$
        \mathbb{P} \left( \frac{n \, \widehat{\V}_b^2(\bX, \, \bY; \, \bZ) }{\widehat{\sigma}_\varepsilon^2} >k \right) = P(T_n > 0),
    $$
where $\widehat{\sigma}_\varepsilon^2 = \frac{1}{n} \sum_{j=1}^n (\widehat{\varepsilon}_j -\tfrac{1}{n} \sum_{i=1}^n \widehat{\varepsilon}_i)^2 $ with
    $$
\widehat{\varepsilon}_i = Y_i - Z_i  (\bZ^t \bZ)^{-1} \bZ^t \bY,
    $$
$T_n$ is defined by
    $$
        T_n = \left(\widehat{\lambda}_1- \frac{k}{n} \right) \, Q_1^2 + \left(\widehat{\lambda}_2- \frac{k}{n} \right) \, Q_2^2 - \frac{k}{n} Q_3^2 - \cdots - \frac{k}{n} Q_{n-p-1}^2
    $$
and $Q_1^2,\ldots,Q_{n-p-1}^2$ are i.i.d. chisquare distributed with one degree of freedom; $\widehat{\lambda}_1$ and $\widehat{\lambda}_2$ are the eigenvalues of the matrix 
    $$
        K = \frac{1}{n} \bU^t (I- {\bZ} ({\bZ}^t {\bZ})^{-1} {\bZ}^t) \bU,
    $$
where $U \in \R^{n \times r}$ is a matrix with entries
    $$
        (\bU)_{ij} = \phi_j(X_i),
    $$
and $\phi_1(X),\ldots, \phi_r(X)$ is an arbitrary feature map of $d_b$.
    
  \end{theorem}

By Proposition \ref{prop:kernel} a feature map with $2$ features exists for each $b \in [0,4]$. Hence $K$ can always be represented by a $2 \times 2$ matrix enabling rapid evaluation of the eigenvalues as demonstrated by the real data example in Section \ref{sec:rd}.
P-values based on Theorem \ref{th:covfinite} can be approximated analogously to the setting without covariates in Section \ref{sec:pvalues}. All results in Section \ref{GT} regarding  the interpretation of $\widehat{\V}_b^2(\bX, \, \bY)$ hold true for $\widehat{\V}_b^2(\bX, \, \bY; \, \bZ)$ with the modification of  adding $\gamma^t Z$ to the right hand side of the corresponding Gaussian regression models.

\section{PRACTICAL ASPECTS} \label{sec:practical}

\subsection{Imputed data}

In practice, GWAS are often performed on imputed genotype data \citep{li2009genotype}. In this case, the SNP information for numerous loci is not directly measured. Instead, the corresponding SNPs are imputed using information from other SNPs and complete data from a reference population. For these imputed SNPs, we do not observe the allele count $X \in \{0,1,2\}$, but the expected allele count $X$ in the interval $[0,2]$. Hence the methodology explained in this paper is not directly applicable in this setting.

A simple but clearly inefficient way to deal with this issue is to round the allele count before performing the analysis. Another straightforward generalization is to use the $\alpha$-distance covariance with $\alpha = \log_2 b$; however this approach leads to a substantially more complicated distribution of test statistic leading to increased computing time.

In order to retain a similar test statistic while using all information on the expected allele counts, we propose to generalize the methodology by linearly interpolating the features, i.e. we use the feature map (cf. Proposition \ref{prop:kernel}),
    $$
    \widetilde{\phi}_1(x) = \sqrt{\frac{b}{2}} x, \quad \widetilde{\phi}_2(x) = \sqrt{\frac{4-b}{2}} |x-1|.
    $$
A straightforward calculation yields that the corresponding distance is,
    $$
d_b(x,y) =  \begin{cases} (x-y)^2 &\quad \text{if } x \geq 1, y \geq 1 \vee x< 1, y < 1, \\ \frac{b}{4} (x-y)^2 +\frac{4-b}{4} (x+y-2)^2 &\quad  \text{if } x \geq 1, y < 1 \vee x< 1, y \geq 1.\end{cases}
    $$
It is easy to see that, Theorem \ref{th:covasy} and \ref{th:covfinite} (which imply Theorems \ref{testasy} and \ref{testfinite} respectively) hold analogously replacing the feature maps in the formulations of the theorems by $(\widetilde{\phi}_1,\widetilde{\phi}_2)^t$.

\subsection{Multiallelic single-nucleotide polymorphisms}

As in most methodological work on GWAS, we were assuming for simplificity that all SNPs are biallelic. However while this assumption is true for the majority of the SNPs, numerous SNPs with three or more alleles (``multi-allelic SNPs'') have been identified \citep{multi}. An advantage of our approach is that it can be straightforwardly generalized to multiallelic SNPs by defining distances on the space $\{0,1,2\}^m$, where $m$ is the number of alleles. We propose the distance,
    $$
    \widetilde{d}_b((x_1,\ldots,x_m),(y_1,\ldots,y_m)) = \frac{1}{2} \sum_{i=1}^m d_b(x_i,y_i),
    $$
where $x_i$ counts the number of alleles of type $i$ and $d_b$ is the distance on $\{0,1,2\}$ used before; this is easily seen to generalize the biallelic case, even in the case of imputed data.

The distribution of the test statistic in the multiallelic setting can be derived similarly as in the biallelic setting, however since there are $m+1 \choose 2$ states the corresponding asymptotic distribution features $t = {m+1 \choose 2} -1$ eigenvalues $\widehat{\lambda_1},\ldots \widehat{\lambda}_t$. If only $2$ alleles are present, the test statistic and its distribution reduce to the biallelic case; very rare alleles have virtually no influence on the test statistic. Since the test is directed towards alternatives corresponding to differences in the most frequent alleles, we expect good power properties in the multi-allelic setting; the test focusses on the variants where there is potentially enough power to detect a possible effect.

\section{SIMULATION STUDY}{\label{sec:simu}}

To demonstrate the performance of our methods and further investigate suitable choices for the parameter $b$, we have conducted a series of simulation studies. Throughout these simulations, we will compare the proposed distance covariance test based on $\widehat{\V}_2$ and $\widehat{\V}_3$ with the following three competitors:
\begin{itemize}
  \item The additive model, performing a linear regression of $y$ on $X$, treating $X \in \{0,1,2\}$ as continuous predictor; this is equivalent  to the test based on $\widehat{\V}_4$ .
  
    \item A linear model treating the SNP $X \in \{0,1,2\}$ as categorical predictor; this model is commonly referred to as ANOVA.
  
    \item A test based on the \texttt{nmax3} statistic, calculated as the maximum of three nonparametric trend tests, based on the recessive, additive and dominant model respectively as implemented in the R package \texttt{AssocTests} \citep{nmax3}
\end{itemize}

\subsection{Computation time}

For evaluating the computation time of the methods, we considered the task of testing the association between $100,000$ SNPs with MAF $0.5$ and a normally distributed response $Y$, which is simulated independent of the SNPs; each method was applied $50$ times - the minimal computation time for each method is displayed. To allow for a fair comparison, we implemented the additive model using our algorithm for $b=4$, i.e. a similar, computationally efficient method is used for the calculation of the test statistic. The ANOVA model is omitted in our comparison, but it can be expected that it shows a very similar performance as the additive model. For our GDC method with $b=2$ and $b=3$, we considered two different versions -  the recommended version described in Section \ref{sec:pvalues} using a screening procedure filtering out SNPs with $p>10^{-3}$ in the first step using a guaranteed anticonservative approximation for our test distribution  (blue and red lines, respectively) and a naive implementation precisely evaluating the p-values for all SNPs (green and yellow lines, respectively). The \texttt{nmax3} procedure (purple line) was implemented using the R package \texttt{AssocTests}. On the left hand side, a comparison of the additive model and the recommended versions for $b=2$ and $b=3$ are provided, highlighting the excellent computational performance of these methods (in particular, $100,000$ SNPs are evaluated in less than $2$ minutes for a sample size of $n=8,000$). On the right side, a comparison of all $6$ methods is given, using a log-scale for the computation time. We note that the naive implementation of the GDC methods without screening leads to a substantially increased computation time, which is more than $10$ times higher than for the version with prescreening. Moreover, it can be noted that the GDC methods without screening show virtually no difference in computation time for the sample sizes under consideration; this is little surprising since the biggest part of the time is used to evaluate the p-values. Finally, we note that the given implementation of the \texttt{nmax3} procedure takes substantially longer computation time than all other methods, making it hard (but not impossible) to apply in practical situations.
 All computations were run on a single core of an Intel Xeon E312xx system with 2.6 MHz.
\begin{figure}
    \centering
    \includegraphics[width=\textwidth]{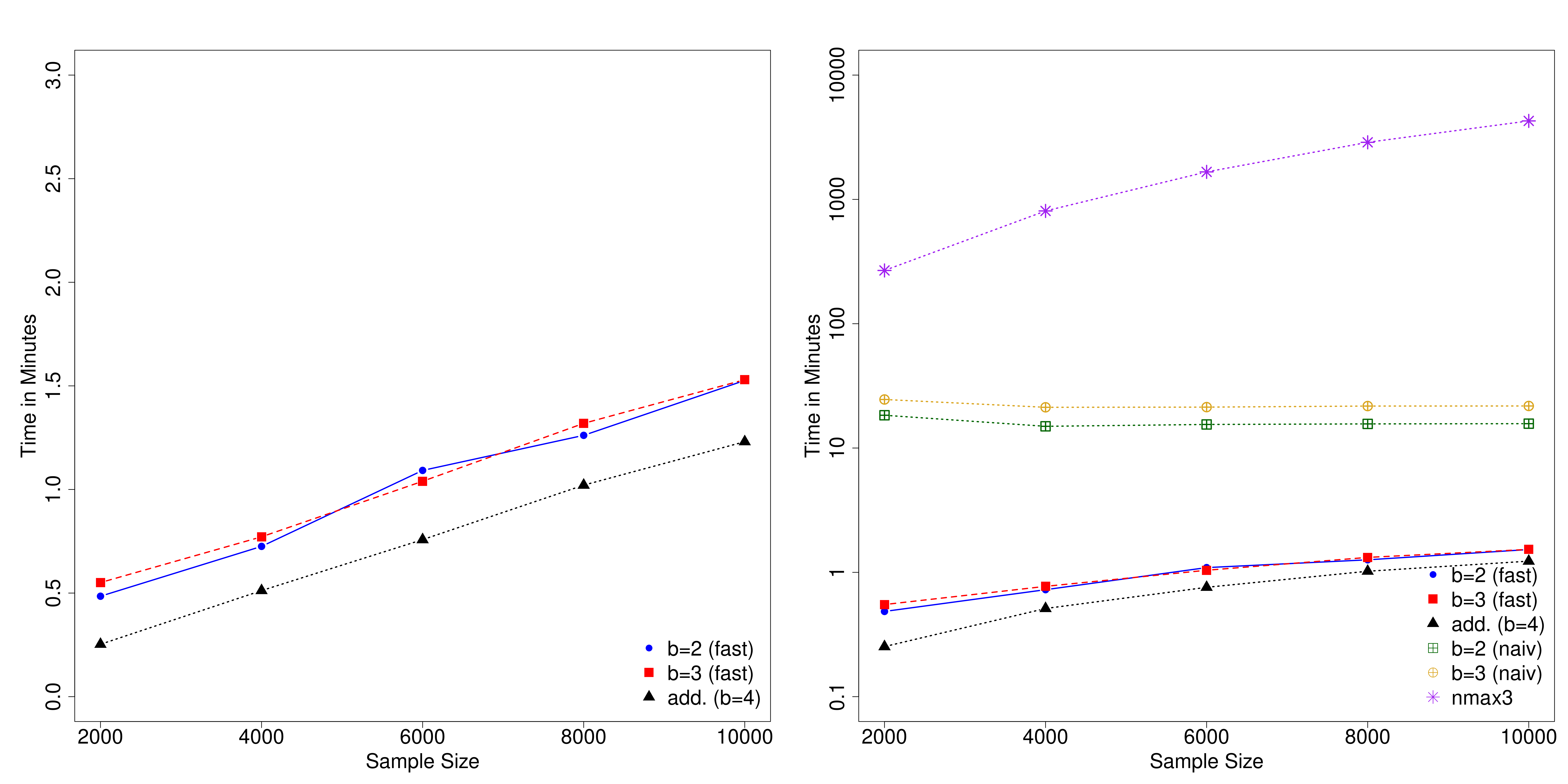}
    \caption{Computation time for different methods for SNP testing on $p=100,000$ SNPs and sample size as indicated in the plot. The different methods are: GDC test with $b=2$ (blue), GDC test with $b=3$ (red), additive model (black), GDC test without p-value screening with $b=2$ (green), GDC test without p-value screening with $b=3$ (yellow) and the \texttt{nmax3} procedure (purple). Left Side: Comparison on Linear Scale. Right Side: Comparison on Log Scale.}.
    \label{fig:het}
\end{figure}

\subsection{Type I error}

For comparing type I error control of the different methods, we fix the sample size at $n=300$ and consider SNPs with MAF $0.1,0.2,\ldots,0.5$.

The data is simulated using a null model under normality, i.e.
   $$
 y_i = \varepsilon_i,
    $$
where $\varepsilon_1,\ldots,\varepsilon_n$ are i.i.d. standard normally distributed random variables. 

\begin{figure}
    \centering
    \includegraphics[width=\textwidth]{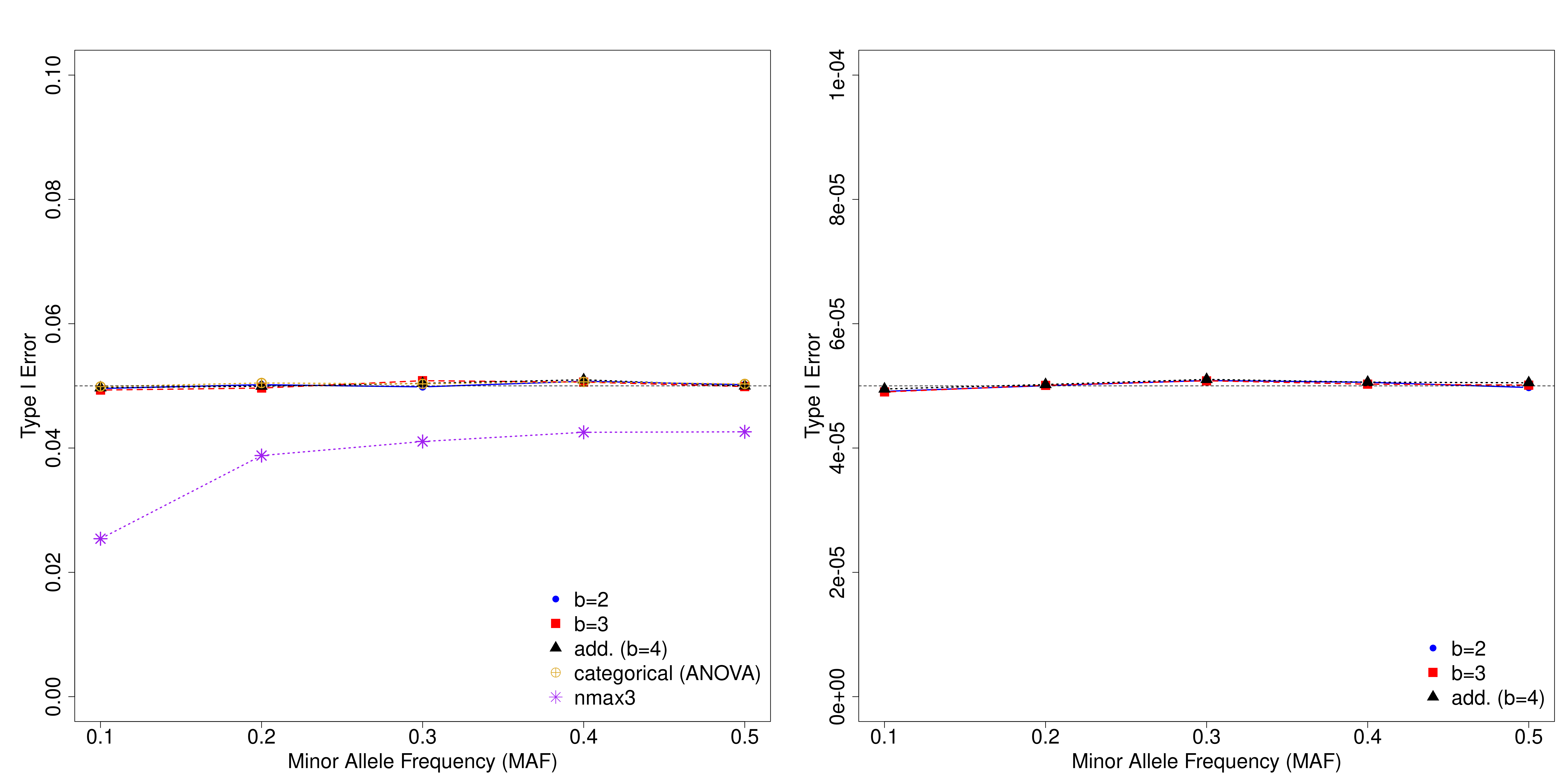}
    \caption{Empirical Type I error rate for different methods for SNP testing for a sample size of $n=300$ and normally distributed outcomes. The left hand-side corresponds to a nominal level of $\alpha = 0.05$ ($10,000$ simulation runs); the right hand-side corresponds to a nominal level of  $\alpha = 5 \times 10^{-5}$ ($100$ million simulation runs). The different methods are: GDC test with $b=2$ (blue), GDC test with $b=3$ (red), additive model (black), ANOVA (yellow) and the \texttt{nmax3} procedure (purple).}.
    \label{fig:typeI}
\end{figure}

We first fix the nominal level at $\alpha = 0.05$ and evaluate the empirical type I error rate for all methods under considerations using  $K=10,000$ simulation runs; the results are provided in the left-hand side of Figure \ref{fig:typeI} . The empirical type I error rate of the tests based on the ANOVA model, the additive model, $\widehat{\V}_2$ and $\widehat{\V}_3$ are always very close to $0.05$. This is not surprising since the exact finite sample distribution is used for all four methods. Our simulation hence confirm Theorem \ref{testfinite}. The \texttt{nmax3} procedure on the other hand is remarkably conservative, particularly for smaller MAF.

To investigate if our methods suffer from numerical issues when approximating the Appell $F_1$ hypergeometric series in \eqref{eq:appell}, we further used $K=100$ million simulation runs to evaluate the empirical type I error rate for a nominal level of $\alpha = 5 \times 10^{-5}$. As can be seen from the right-hand side of Figure \ref{fig:typeI}, the empirical type I error rate  of our methods is again very close to the nominal level.

A type I error rate approximation for the genome-wide significance level $\alpha = 5 \times 10^{-8}$ is omitted; to achieve a similar precision as for $\alpha = 5 \times 10^{-5}$, ca. $K=100$ billion simulations would be required. 

\subsection{Power}

For comparing the power of the different methods, we assume the model,
\begin{equation} \label{eq:powermodel}
 y_i = h \beta \, 1_{\{X=1\}} +  \beta \, 1_{\{X=2\}} + \varepsilon,
\end{equation}
where $\varepsilon$ follows a normal distribution with mean $0$ and variance $25$. We let the heterozygous effect $h$ vary with $h= 0,0.1,\ldots,1$ This can be considered an approximation of a situation, where the quantitative trait is the sum of $26$ independent terms of the same size and we consider the power for detecting one of the terms.

We consider two situations; in the first the sample size is $n=300$ and the nominal level is $\alpha = 0.05$, for the second the sample size is $n=3000$ and $\alpha = 5 \times 10^{-8}$. For the MAF, we choose $0.1$, $0.25$, $0.4$ and $0.5$; for the $n=3000$ setting, we additionally consider two scenarios with rare alleles  (MAFs $0.05$ and $0.01$) and increased sample sizes ($n = 6,000$ and $n=30,000$, respectively). 

\begin{figure}
    \centering
    \includegraphics[width=\textwidth]{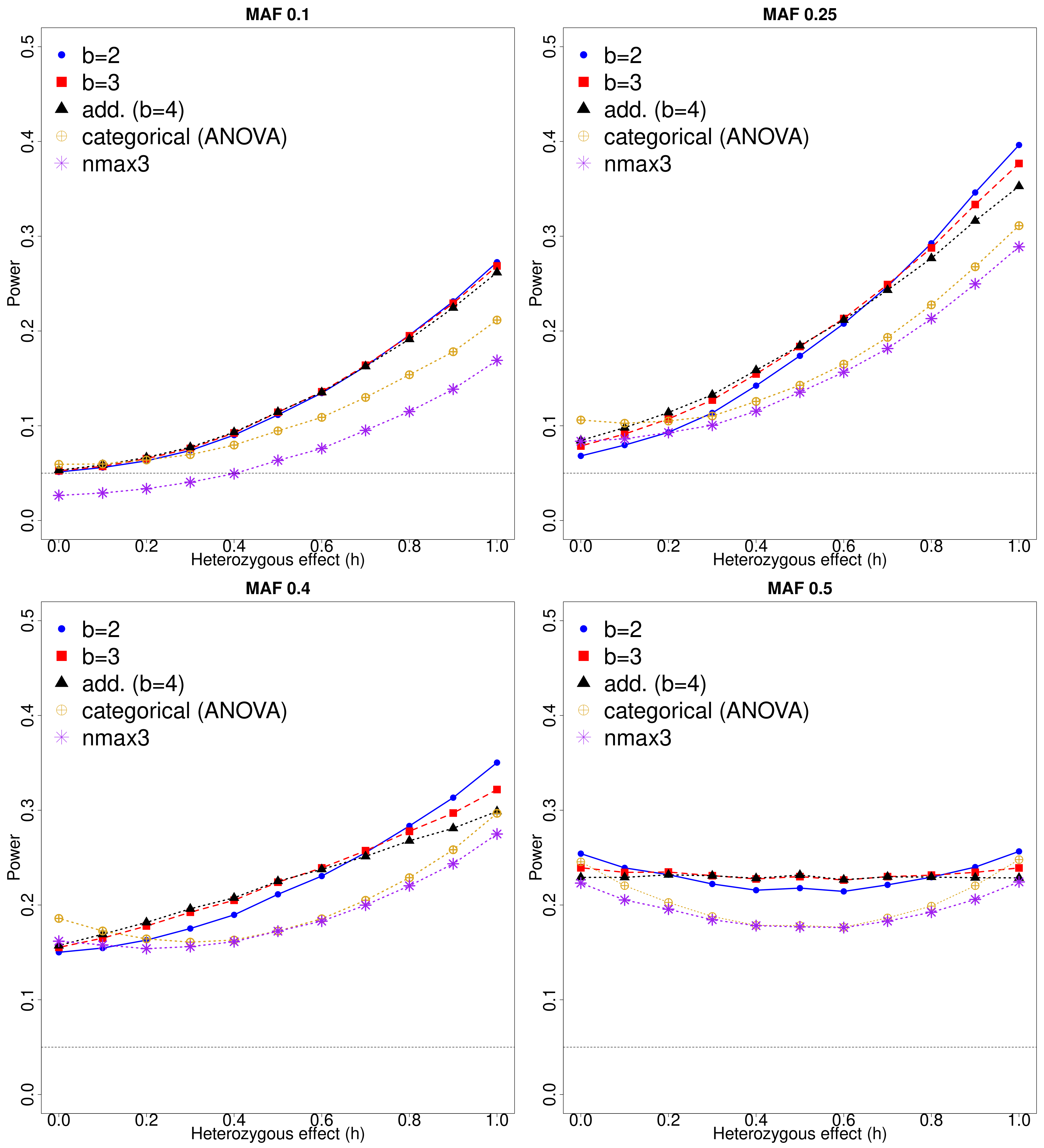}
    \caption{Power for different methods for SNP testing for a sample size of $n=300$ and normally distributed outcomes using a nominal level of $\alpha =0.05$. We follow the model \eqref{eq:powermodel} with heterozygous effect $h$ as indicated and MAF of $0.1$ (left plot), $0.3$ (middle plot) and $0.5$ (right plot). The different methods are: GDC test with $b=2$ (blue), GDC test with $b=3$ (red), additive model (black), ANOVA (yellow) and the \texttt{nmax3} procedure (purple).}.
    \label{fig:power05}
\end{figure}

Figure \ref{fig:power05} illustrates the results for the nominal level $0.05$ and $n=300$. For MAF $0.5$, the proposed GDC test with $b=2$ dominates the ANOVA approach and the \texttt{nmax3} procedure for all considered models. The difference between the GDC test for $b=3$ and the additive model (which is equivalent to the GDC test with $b=4$) are rather small, with tiny (nonvisible) advantages for the additive model for $h \in \{0.4,0.5,0.6\}$ and slight advantages for $b=3$ for all other values of $h$ (e.g. power $0.239$ vs. $0.229$ for $h=0$). The power for the GDC test with $b=2$ is greater than the tests for $b=3$ and $b=4$ for dominant recessive models ($h \in \{0,1\}$, but smaller for values of $h$ that reflects models that are close to additive.

For smaller MAFs, we make the observation that the ANOVA approach typically has the best performance for the recessive model, but shows substantially worse performance than the GDC tests for other values. On the other hand, the ANOVA approach dominates the \texttt{nmax3} procedure in all settings. For the different GDC tests, we note that $b=4$ typically gives the best performance for the recessive model, while $b=2$ is best for the dominant model. With decreasing MAF, the power curves of the three GDC tests get closer and are nearly indistinguishable for MAF $0.1$.

The power simulation for MAF $0.3$ also provides a practical confirmation of the locally most powerful properties studied in Section \ref{GT}. Notably, the average of the power for the recessive ($h=0$) and dominant ($h=1$) is always highest for the GDC test with $b=2$ (cf.  left plot in Figure \ref{fig:lomopo}). Moreover a simple trapezoidal integration (not shown) of the power curves yields that the GDC test with $b=3$ has  the highest integrated power in all settings, indicating that this test is best for settings where $h$ is uniformly distributed (cf. middle plot in Figure \ref{fig:lomopo}. Finally the GDC test with $b=4$ is always best for the additive model (cf. right plot in Figure \ref{fig:lomopo}).

\begin{figure}
    \centering
    \includegraphics[width=\textwidth]{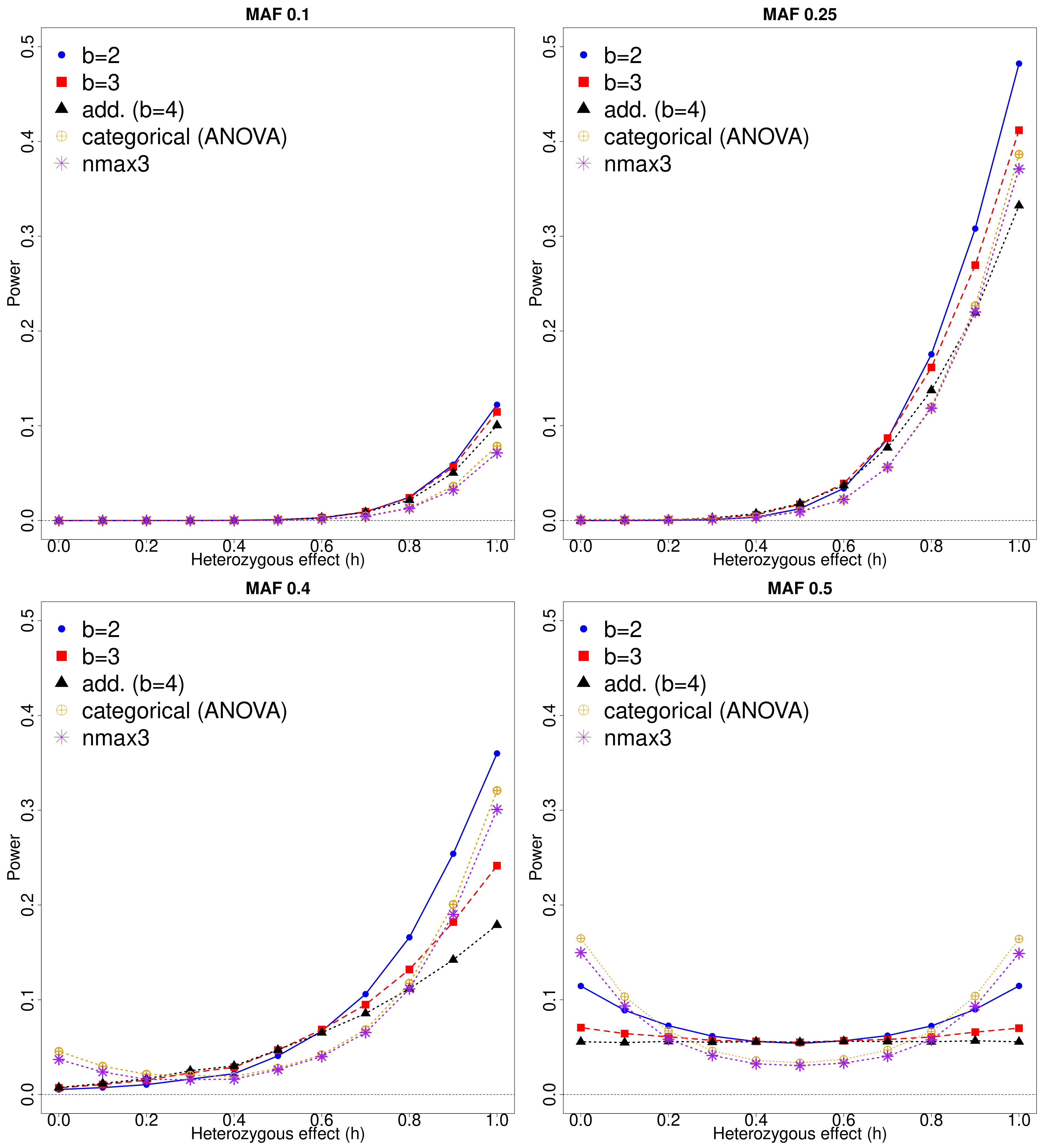}
    \caption{Power for different methods for SNP testing for a sample size of $n=3,000$ and normally distributed outcomes using a nominal level of $\alpha = 5 \times 10^{-8}$. We follow the model \eqref{eq:powermodel} with heterozygous effect $h$ and MAF as indicated. The different methods are: GDC test with $b=2$ (blue), GDC test with $b=3$ (red), additive model (black), ANOVA (yellow) and the \texttt{nmax3} procedure (purple).}.
    \label{fig:power5e8}
\end{figure}

The results of the power simulation for $n=3,000$ and $\alpha = 5 \times 10^{-8}$ given in Figure \ref{fig:power5e8} differ substantially from the first set of power simulations. A possible explanation is that due to the increased sample size, the effect is in a certain sense greater or ``less local''. For MAF $0.5$, we now observe that the ANOVA model yields the best power for $h \in \{0,0.1,0.9,1\}$. The power curves for the different GDC tests are nearly indistinguishable for values of $h$ reflecting a model that is close to additive $h \in \{0.4,0.5,0.6\}$, but show moderate advantages for $b=3$ and clear advantages for $b=2$ for other values. Interestingly - in our simulations for MAF $0.5$ - the empirical power of the test for $b=3$ was always at least as high as the power for $b=4$ (equal for $h=0.5$ and better for all other values). For smaller MAF, we see a similar trend as described for the first set of simulations, with clear advantages of $b=2$ and $b=3$ compared to $b=4$ for models that are close to dominant and slight to very slight disadvantages for some other models. For $MAF$ 0.1, the three different GDC test are very close.

\begin{figure}
    \centering
    \includegraphics[width=\textwidth]{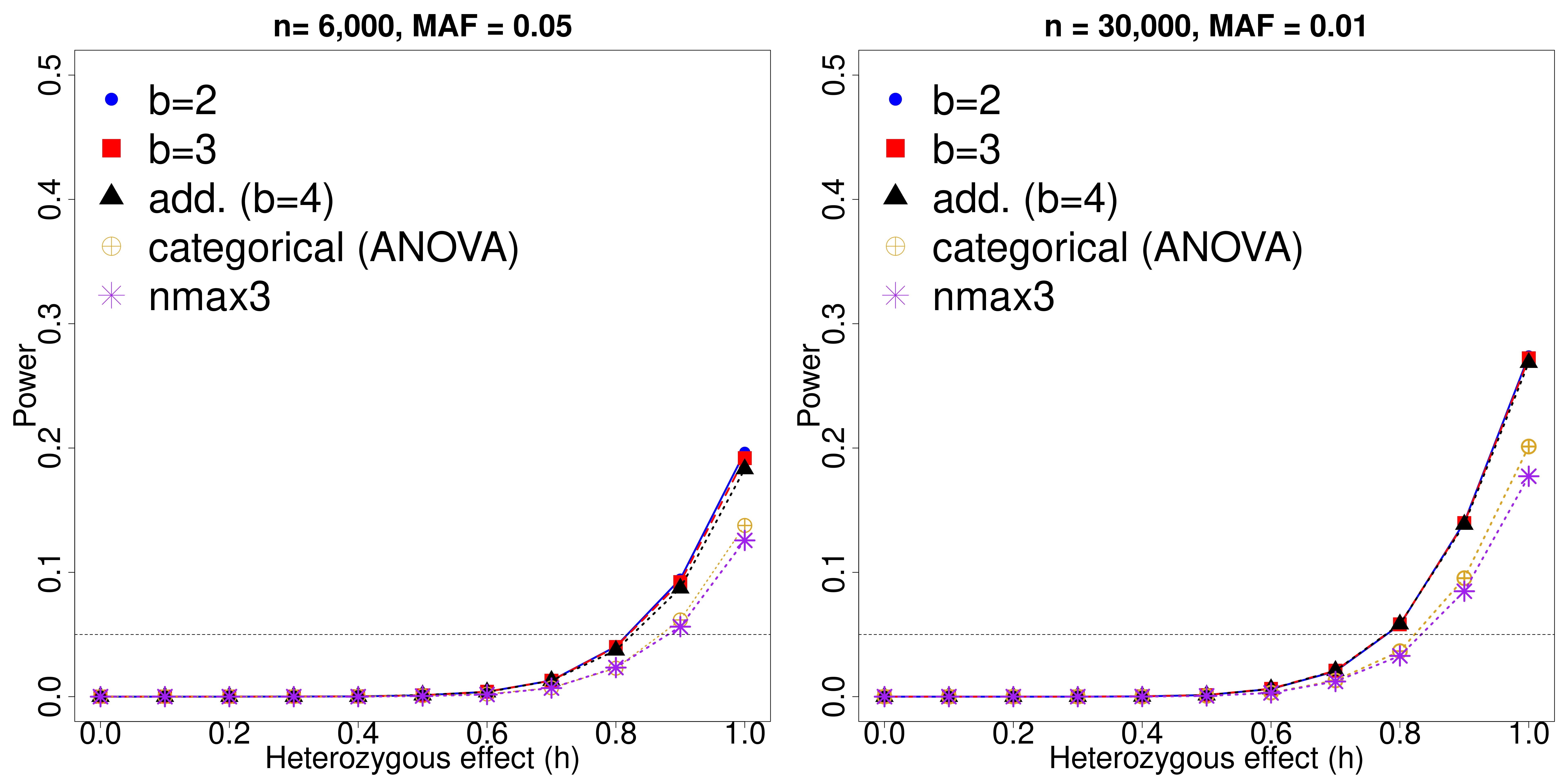}
    \caption{Power for different methods for SNP testing for a sample size as indicated and normally distributed outcomes using a nominal level of $\alpha = 5 \times 10^{-8}$. We follow the model \eqref{eq:powermodel} with heterozygous effect $h$ and MAF as indicated. The different methods are: GDC test with $b=2$ (blue), GDC test with $b=3$ (red), additive model (black), ANOVA (yellow) and the \texttt{nmax3} procedure (purple).}.
    \label{fig:rare}
\end{figure}

Figure \ref{fig:rare} provides the results of the power simulations for the rare SNPs, illustrates that the performance of the three GDC tests is essentially the same for these settings, while the GDC tests clear dominate the ANOVA test and \texttt{nmax3} proecedure.

\section{REAL DATA ANALYSIS} \label{sec:rd}

We now show how the proposed methodology can be applied to real data. One of the interests of psychiatric genetics is the study of addictions \citep{Hatoum}, with special focus in substance use disorders such as alcoholism \citep{Shield}. 
Large-scale analyses are starting to reveal the polygenic architecture of several alcohol-related traits \citep{Gelernter}. We will focus on cirrhosis, a form of hepatic disease which is modulated by some prominent liver enzymes such as aspartate aminotransferase (AST), alanine aminotransferase (ALT) and $\gamma$-glutamyltransferase (GGT). The search for new loci associated with liver enzymes is a topic of current research interest \citep{Ghouse,Pazoki}. 

\begin{figure}[!htbp]
	\centering\includegraphics[width=0.8\textwidth]{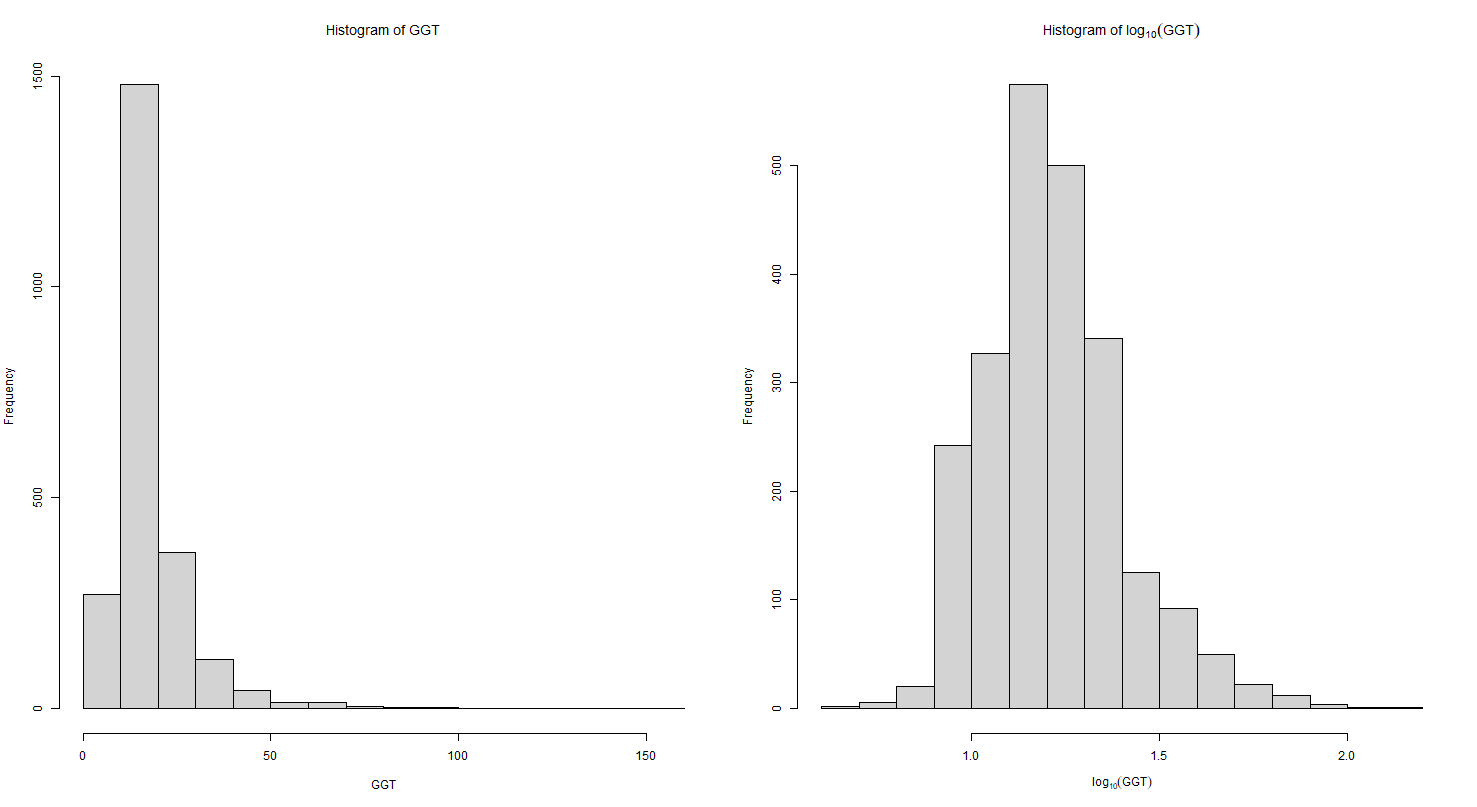}
	\caption{Histograms of the raw values of the GGT serum concentration for the Trinity data (left) and its logarithm with base $10$.}
	\label{hist:ggt}
\end{figure}

The dataset is derived from the Trinity Student Study (dbGaP accession number: \href{https://www.ncbi.nlm.nih.gov/projects/gap/cgi-bin/study.cgi?study_id=phs000789.v1.p1}{phs000789.v1.p1}) and has been described in \citet{Mills:Molloy,Molloy:Brody:2016,Desch}. The cohort was sampled during the academic year 2003--2004 in the Trinity College of the University of Dublin, with the goal of researching the genetics of quantitative traits. Only students with no serious medical condition, and of Irish ethnicity (based on the geographic origin of their grandparents), were included. Further basic descriptive statistics of the dataset are displayed in Table~\ref{data:descr:table}. The phenotype under consideration is the GGT serum concentration, which we log-transformed for the purposes of our analysis, see Figure \ref{hist:ggt}. All $757,577$ SNPs in the dataset were used for association testing.

\begin{table}[!htbp]
	\centering
	\caption{Descriptive statistics for the Trinity Student Study dataset}
\begin{tabular}{l|l}
	\hline
	Sample size & $N=2407$ individuals \\
	Age & median $22$, range $[19, 28]$ \\
	Sex & $1409 \, (58.5 \%)$ females and $998 \, (41.5 \%)$ males \\
	GGT serum concentration &  median $15$, range $[5,158]$ \\
	\hline
\end{tabular}
\label{data:descr:table}
\end{table}

\begin{figure}[!htbp]
	\centering\includegraphics[width=0.98\textwidth]{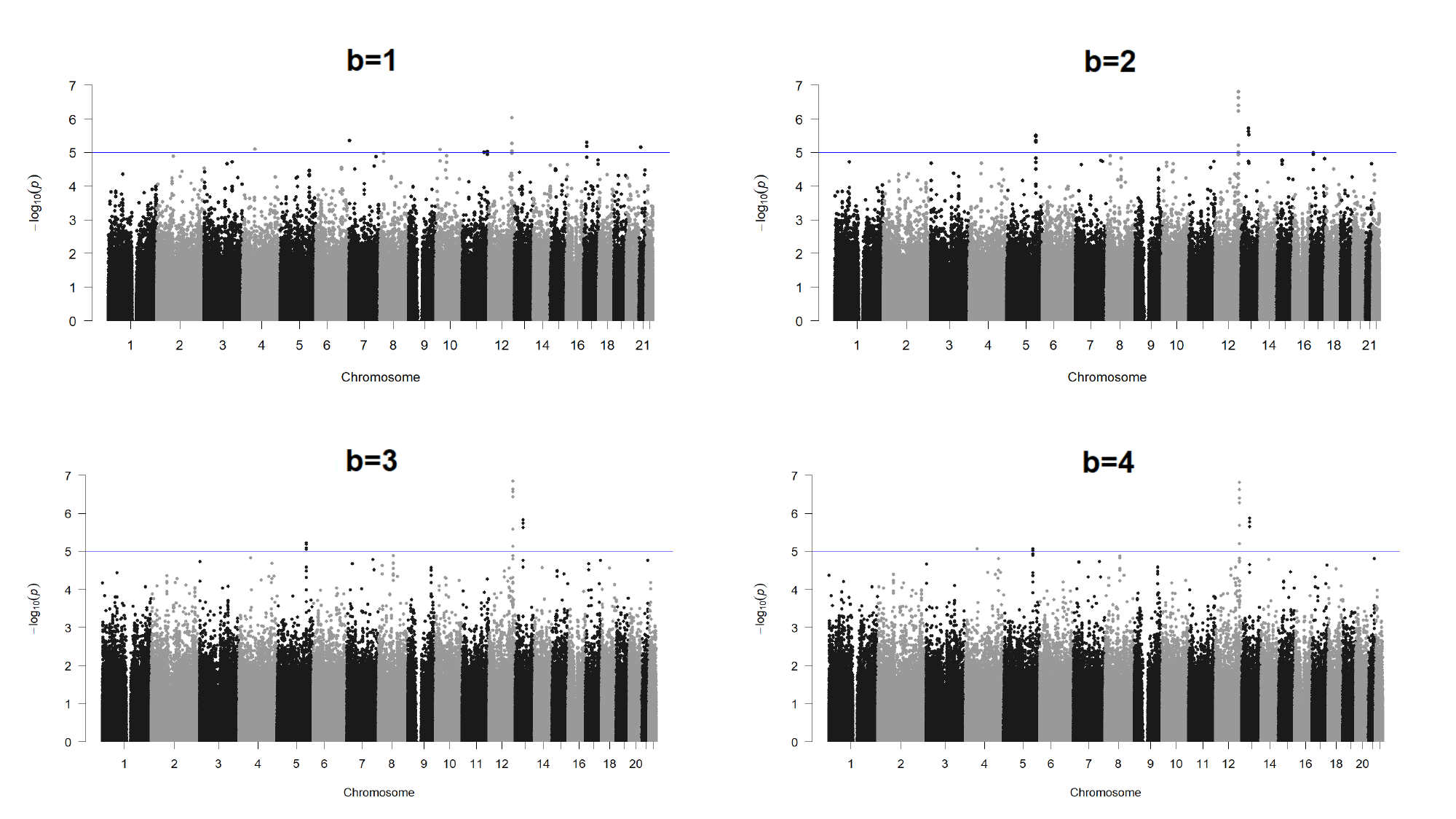}
	\caption{Manhattan plots for the Trinity dataset analysis using the the proposed GDC test, for $b\in\{1,2,3,4\}$ linearly adjusting for age and sex. Blue horizontal lines indicate a significance threshold of $10^{-5}$. The GDC test with $b=4$ corresponds to the standard linear regression test based on an additive}
	\label{manh_dc_agesx}
\end{figure}

We have applied the method described in Theorem \ref{th:covfinite} to test for the association of all $757,577$ SNPs  with log GGT serum concentration linearly adjusting for the covariates age and sex. For the parameter $b$, the values $1$, $2$, $3$ and $4$ were used.  The results are displayed in Figure~\ref{manh_dc_agesx} as Manhattan plots. These graphics are a standard visualisation in GWA studies plotting minus $\log_{10}$ of the $p$-value against the physical location of each SNP considered in the genome (left to right, chromosomes 1 to 22; and then ordering within them according to the nucleotide position). While no associations passed the genome-wide significance threshold of $5 \times 10^{-8}$, several SNPs showed an association with a p-value smaller than $10^{-5}$ ($12$ SNPs for $b=1$, $15$ SNPs for $b=2$, $16$ SNPs for $b=3$ and $13$ SNPs for $b=4$). The highest ``skyscrapers'' for each of the four different methods are located on chromosome 12, in the region of the hepatocyte nuclear factor 1 homeobox A (HNF1A) gene. HNF1A encodes the protein hepatocyte nuclear factor-1 alpha (HNF-1$\alpha$), which is required for the expression of several liver-specific genes \cite{hnf11}. The most notable association is the one with rs1169288, a missense mutation in HNF1A that has been previously shown to be associated with GGT in an Australian cohort \cite{Middelberg}. The p-value for the association of GGT and rs1169288 was smaller than $10^{-5}$ for all values of $b$ ($9.2 \times 10^{-7}$ for $b=1$, $1.6 \times 10^{-7}$ for $b=2$, $1.4 \times 10^{-7}$ for $b=3$ and $1.5 \times 10^{-7}$ for $b=4$). 



\section{DISCUSSION}\label{Discu}


In this work, we have derived novel methodology for testing the association of a single SNP with a quantitative response based on a version of generalized distance covariance. The test is computationally efficient, has good statistical properties and can be applied with nuisance covariates. For each value of the parameter $b$, the corresponding test can be interpreted a the locally most powerful test in certain situations. Simulation studies indicate that our approach indeed offers a powerful alternative to the predominant additive model.

An important limitation of the provided methodology is that it is currently restricted to quantitative endpoints and does not allow for survival endpoints or binary data. Moreover, the approach is constrained to single SNPs, while in practice also testing for the association of a group of SNPs (for example in the region of a gene) with a response is interesting. The extension of the methodology to survival endpoints, GLMs and groups of SNPs is subject of current and future research.

As another extension, we aim at generalizing the methodology to testing with general ordinal data. The results in Section \ref{GT} (cf. \ref{fig:lomopo}) show that, for $b \in [2,4]$, the corresponding test represents a sensible approach for dealing with the association of anordinal variable with cardinality $3$ and a quantitative response. It hence appears natural to consider extensions for ordinal variables with more than three states.

A particularity of our approach - that can be considered either a drawback or an advantage - is that each value of $b \in [0,4]$ offers a different test, which makes it necessary that the practitioner makes a choice about this parameter before applying the testing procedure. We recommend $b=3$ as a preliminary default parameter. While Figure \ref{fig:power5e8} shows that $b=2$ offers higher power than all other methods in many scenarios, we could identify barely any scenario, where $b=3$ performs considerably worse than the additive model. Recognizing that more experience with real data is needed to make a final decision on this issue, we have a preference for $b=3$, since this puts us in little danger to do worse than the standard methodology.  

\section*{ACKNOWLEDGEMENTS}


This work has been supported by project \mbox{PID2020-116587GB-I00}, granted by {MICIU/AEI/10.13039/ 501100011033} (Spanish Ministry of Science).

We thank the participants of the Trinity Student Study (dbGaP accession \textit{phs000789.v1.p1}), first published by \citet{Mills:Molloy}. This study was supported by the Intramural Research Programs of the National Institutes of Health, the National Human Genome Research Institute, and the Eunice Kennedy Shriver National Institute of Child Health and Development.

We also acknowledge the valuable advice received by Dr Javier Costas with regard to the real data example, and the help of Prof Rosa Crujeiras in accessing dbGaP.

FCP is grateful to the the DKFZ Biostatistics Unit for all their support before, during and after his research stay in Heidelberg, and more generally to each person who contributed to make those months so beautiful.

\section{DATA AVAILABILITY STATEMENT}

Access to the data of the Trinity Student Study can be requested via dbGaP (\url{https://www.ncbi.nlm.nih.gov/projects/gap/cgi-bin/study.cgi?study_id=phs000789.v1.p1}).

\vspace*{.25cm}

\section*{REFERENCES}

\settocbibname{}

\bibhang 0em

\end{document}


\maketitle

\appendix

\section{Proofs of the theoretical results}

\Pro of Theorem 3.1.  By  \cite[Eq. (4)]{DJ}, the generalized distance covariance $\V_{\rho_\mX, \rho_\mY}$ can be written as 
\begin{equation} \label{eq:dcovcov2}
    \V^2_{\rho_\mX, \rho_\mY}(X,Y) = \sum_{l=1}^{d_\mX} \sum_{m=1}^{d_\mY} \mbox{Cov}^2 (\phi_l^{\rho_\mX} (X), \phi_m^{\rho_\mY} (Y)),
\end{equation}
where $(\phi_1^{\rho_\mX}, \ldots, \phi_{d_\mX}^{\rho_\mX})$ and $(\phi_1^{\rho_\mY}, \ldots, \phi_{d_\mY}^{\rho_\mY})$ are feature maps of the premetrics $\rho_\mX$ and $\rho_\mY$, respectively. \\
The premetric $\rho_\mY (y,y') = \tfrac12 |y-y'|^2$ induces the linear kernel $l(y,y) = y y'$ with trivial feature map $\phi^{\rho_\mY}  = id$, a feature map of $d_b$ is given by (cf. Proposition 4.2), 
$$
\phi_1^{d_b}(x) = \sqrt{\frac{b}{2}} (-1_{\{x=0\}} + 1_{\{x=2\}}) , \quad \phi^{d_b}_2(x) = \sqrt{\frac{4-b}{2}} 1_{\{x=1\}}.
$$
Inserting these feature maps into Eq. \eqref{eq:dcovcov2}, we obtain,
$$
\V_b^2(X,Y) = \frac{b}{2} (\mbox{Cov}(-1_{\{X=0\}} + 1_{\{X=2\}}, \, Y))^2 \, +  \, \frac{4-b}{2} (\mbox{Cov}(1_{\{X=1\}}, \, Y))^2.
$$
Using elementary transformations and applying the law of total probability, we obtain
$$
\V_b^2(X,Y) =  \frac{b}{2} \, (- p_0 \, (\mu_0 - \mu_Y) +  p_2 \, (\mu_2 - \mu_Y))^2 + \frac{4-b}{2} \, (p_1 \, (\mu_1 - \mu_Y))^2,
$$
where $\mu_Y = E[Y]$.

If $\mu_0 = \mu_1 = \mu_2 = \mu_Y$, it  follows that $\V_b(X,Y) = 0$, completing the proof of the first part.


For the second part, assume that $\mu_i \neq \mu_j$ for some $i \neq j$. We first take care of the case $\mu_1 \neq \mu_Y$. Then 
    $$
    \frac{4-b}{2} \, (p_1 \, (\mu_1 - \mu_Y))^2 > 0,
    $$
and hence $\V_b(X,Y)>0$. 

Now consider the remaining case $\mu_1 = \mu_Y$. In this case, either $\mu_0 < \mu_Y < \mu_2$ or 
$\mu_2 < \mu_Y < \mu_0$. For either possibility, it follows that
    $$
    \frac{b}{2} \, (- p_0 \, (\mu_0 - \mu_Y) +  p_2 \, (\mu_2 - \mu_Y))^2 > 0,
    $$
and hence 
$\V_b(X,Y)>0$.
\qed

\bigskip

\Pro of Proposition 3.2. 

We consider first $b=0$. For any $(X,Y) \in \{0,1,2\} \times \R$, by the law of total probability,
    $$
    \mu_Y = p_0 \mu_0 + p_1 \mu_1 + p_2 \mu_2.
    $$
Let $Y$ be such $\mu_0 = p_2, \mu_1 = 0$ and $\mu_2 = - p_0$.
Then $\mu_0 \neq \mu_2$, but $\mu_1 = \mu_Y$ and hence,
    $$
    \V_0(X,Y) = 2 \, (p_1 \, (\mu_1 - \mu_Y))^2 = 0.
    $$
    
Now consider $b=4$.
Let $Y$ be such $\mu_0 = p_1 p_2$, $\mu_1 = - 2 p_0 p_2$ and $\mu_2 = p_0 p_1$. Then $\mu_0 \neq \mu_1$, however $\mu_Y = 0$ by the law of total probability and hence

    $$
    \V_4(X,Y) = 2 \, (- p_0 \, (\mu_0 - \mu_Y) +  p_2 \, (\mu_2 - \mu_Y))^2 = 0.
    $$
\qed

\bigskip

\Pro of Theorem 3.3. As in the proof of Theorem 3.1, we consider the feature map corresponding to $d_b$ given by the vector notation,
    $$
        \phi_1 = \sqrt{\frac{b}{2}} \begin{pmatrix} -1 \\ 0 \\1 \end{pmatrix}, \, \phi_2 = \sqrt{\frac{4-b}{2}} \begin{pmatrix} 0 \\ 1 \\ 0 \end{pmatrix},
    $$
i.e.
$$
\phi_1(x) = \sqrt{\frac{b}{2}} (-1_{\{x=0\}} + 1_{\{x=2\}}) , \quad \phi_2(x) = \sqrt{\frac{4-b}{2}} 1_{\{x=1\}}.
$$
We further define $U_1 = \phi_1(X)$, $U_2 = \phi_2(X)$, $\mu_U = (E[U_1], E[U_2])^t$ and $\mu_Y = E[Y]$. Finally, for a sample of size $n$, for each $i \in \{1,\ldots,n$\}, $U_{1i} = \phi_1(X_i)$, $U_{2i} = \phi_2(X_i)$ and $\bU$ is the corresponding data matrix in $\mathbb{R}^{n \times 2}$ with
    $$
        (\bU)_{kl} = U_{lk}, \quad k \in \{1,\ldots,n\},\, l \in \{1,2\}.
    $$
Finally let $I$ denote the $n \times n$ identity matrix, $\mathbf{1} = (1,\ldots,1)^t \in \mathbb{R}^n$ and $H=\frac{1}{n} \mathbf{1} \mathbf{1}^t$.
From \cite[Eq. (3)]{DJ}, it follows that $\widehat{\V}_b^2(\bX,\bY)$ can be written as
    \begin{align*}
n \, \V_b^2(\bX,\bY) &= \frac{1}{n} \bY^t (I-H) \bU \bU^t (I-H) \bY \\ &= \bv^t \bv,
\end{align*}
with 
$$
\bv = \frac{1}{\sqrt{n}} \bU^t (I-H) \bY.
$$
Since $(I-H) \ba = 0$ for any vector with constant components $\ba=a\mathbf{1}$, $\bv$ can alternatively be written as
\begin{align}
\bv &= \frac{1}{\sqrt{n}} (\bU- \mathbf{1} \mu_U^t)^t (I-H) (\bY-\mathbf{1} \mu_Y) \\
&= \frac{1}{\sqrt{n}} (\bU- \mathbf{1} \mu_U^t)^t (\bY-\mathbf{1} \mu_Y) - \frac{1}{\sqrt{n}} (\bU- \mathbf{1} \mu_U^t)^t H (\bY-\mathbf{1} \mu_Y) \label{eq:vdecomp} .
\end{align}
We first consider the second term in Equation \eqref{eq:vdecomp},
\begin{align*}
    \frac{1}{\sqrt{n}} (\bU- \mathbf{1} \mu_U^t)^t H (\bY- \mathbf{1} \mu_Y) &= 
    \frac{1}{n^{3/2}} (\bU- \mathbf{1} \mu_U^t)^t \mathbf{1} \mathbf{1}^t (\bY-\mathbf{1} \mu_Y) \\ &= \frac{1}{n^{3/2}} \begin{pmatrix}
\sum_{i=1}^n \left(U_{1i} - E[U_1]\right) \\
\sum_{i=1}^n \left(U_{2i} - E[U_2]\right)
\end{pmatrix} \left(\sum_{i=1}^n \left(Y_{i} - E[Y]\right)\right). 
\end{align*}
Since $\frac{1}{\sqrt{n}} \begin{pmatrix}
\sum_{i=1}^n \left(U_{1i} - E[U_1]\right) \\
\sum_{i=1}^n \left(U_{2i} - E[U_2]\right)
\end{pmatrix} $ and $\frac{1}{\sqrt{n}} \sum_{i=1}^n \left(Y_{i} - E[Y]\right)$ converge in probability to normal distributions due to the multivariate central limit theorem (CLT), this term converges in probability to zero.
We now consider the first term in Equation \eqref{eq:vdecomp}. Applying the multivariate CLT and using that $U$ and $Y$ independent, we observe that, for $n \to \infty$
$$
\frac{1}{\sqrt{n}} (\bU- \mathbf{1} \mu_U^t)^t (\bY-\mathbf{1} \mu_Y) \stackrel{\mathcal{D}}{\longrightarrow} N(0,\Gamma),
$$
where 
    $$
\Gamma = \begin{pmatrix} \sigma_Y^2 \, Var(\phi_1(X)) & \sigma_Y^2\, Cov(\phi_1(X), \phi_2(X)) \\ \sigma_Y^2 \, Cov(\phi_1(X), \phi_2(X)) &  \sigma_Y^2 \, Var(\phi_2(X)) \end{pmatrix}    
    $$
We will assume in the following that $\Gamma$ has full rank; in cases where the rank of $\Gamma$ equals one, the proofs can be carried out similarly. Then
\begin{equation} \label{eq:wconv}
        \bw := \Gamma^{-1/2} \bv \stackrel{\mathcal{D}}{\longrightarrow} N(0, I_2)
\end{equation}
where $I_2$ is the $2 \times 2$ identity matrix. Consequently, we obtain 
\begin{align}
    n \V_b^2(\bX,\bY) &= \bw^t \,  \Gamma \, \bw  \nonumber \\ &= \sigma_Y^2 \, \bw^t \, Q \, \Lambda \, Q^t \, \bw, \label{eq:dcovfinalform}
\end{align}
where $Q$ is an orthogonal $2 \times 2$ matrix, $\Lambda$ is a diagonal matrix of the form 
   $$
\Lambda = \begin{pmatrix} \lambda_1 & 0  \\ 0 &  \lambda_2 \end{pmatrix},    
    $$
and $\lambda_1, \lambda_2$ are the eigenvalues of the matrix 
 $$
 K = \begin{pmatrix}  Var(\phi_1(X)) &  Cov(\phi_1(X), \phi_2(X)) \\ Cov(\phi_1(X), \phi_2(X)) &   Var(\phi_2(X)) \end{pmatrix} 
 $$
Evaluation of the entries of $K$ is straightforward and yields the form given in the main manuscript.
Since the standard normal is invariant under orthogonal transformations, combining Equations \eqref{eq:wconv} and \eqref{eq:dcovfinalform} yields
    $$
   n \V_b^2(\bX,\bY)  \stackrel{\mathcal{D}}{\longrightarrow} \sigma_Y^2 (\lambda_1 Q_1^2 + \lambda_2 Q_2^2),
    $$
where $Q_1^2$ and $Q_2^2$ are chisquare distributed. This completes the proof.
\qed

\bigskip

\Pro of Theorem 3.4. We use the same notation as in the proof of Theorem 3.3. Hence
$\widehat{V}_b^2(\bX,\bY)$ can again be written as
    \begin{align*}
n \, \V_b^2(\bX,\bY) &= \frac{1}{n} \bY^t (I_n-H) \bU \bU^t (I_n-H) \bY \\ &= \frac{1}{n}  (\bY-1 \mu_Y)^t (I_n-H) \bU \bU^t (I_n-H) (\bY-1 \mu_Y), 
\end{align*}
where the second line follows from $1 \mu_Y = H 1 \mu_Y$.
Consequently,
    $$
 \frac{n \, \widehat{\V}_b^2(\bX,\bY)}{\widehat{\sigma}_Y^2}  =  \frac{(\bY-1 \mu_Y)^t (I_n-H) \bU \bU^t (I_n-H) (\bY-1 \mu_Y)}{(\bY-1 \mu_Y)^t (I_n-H) I_n (I_n-H) (\bY-1 \mu_Y)}.
    $$
Hence
$$
 \left\{ \frac{n \, \widehat{\V}_b^2(\bX,\bY)}{\widehat{\sigma}_Y^2} \geq k \right\}
$$
is obviously equivalent to
$$
 \left\{ \frac{1}{n} (\bY-1 \mu_Y)^t (I_n-H)  (\bU \bU^t - k I_n) (I_n-H) (\bY-1 \mu_Y) \geq 0 \right\}.
$$
Now consider the matrix
\begin{align*}
 \frac{1}{n} (I_n-H) (\bU \bU^t  - k I_n) (I_n-H) =  \frac{1}{n} (I_n-H) \bU \bU^t   (I_n-H) - \frac{k}{n} \, (I_n - H).
 \end{align*}
 The constant vector $\bo_n = \left(\sqrt{\frac{1}{n}}, \ldots,\sqrt{\frac{1}{n}}\right)^t$ is an eigenvector to eigenvalue $0$ for both the matrices $(I_n-H) \bU \bU^t   (I_n-H)$ and $(I_n - H)$. Augmenting $\bo_n$ to an orthogonal basis (represented by the matrix $O$) of $(I_n-H) \bU \bU^t   (I_n-H)$, we obtain,
 \begin{align*}
\frac{1}{n}  (I_n-H) \bU \bU^t   (I_n-H) - \frac{k}{n} \, (I_n - H) 
 &= O \widehat{\Lambda} O^t -  O  D_{n-1} O^t,
\end{align*}
where $D_{n-1}$ is a diagonal matrix with diagonal $(k/n,k/n,\ldots,k/n,0)$. Since the standard normal distribution is invariant under orthogonal transformations, we obtain that
\begin{align*}
&(\bY-1 \mu_Y)^t (I_n-H) (\bU \bU^t - k I_n) (I_n-H) (\bY-1 \mu_Y) \\ &\stackrel{\mathcal{D}}{=} (\widehat{\lambda}_1 - k/n) Q_1^2 + (\widehat{\lambda}_2 - k/n) Q_2^2 - k/n Q_3^2 - \cdots - k/n Q_{n-1}^2,
\end{align*}
where $Q_1^2, \ldots, Q_{n-1}^2$ are chisquare distributed and $\widehat{\lambda}_1$ and $\widehat{\lambda}_2$ are the eigenvalues of $\widehat{K} = \frac{1}{n}  \bU^t (I_n-H) \bU $. The evaluation of the entries in $\widehat{K}$ is straightforward and completes the proof.
\qed

\bigskip

\Pro of Proposition 3.5. Throughout this proof, we will use the usual stochastic ordering  \citep{Shaked},
    $$
X \leq_{st} Y \Leftrightarrow \mathbb{P}(X > c) \leq \mathbb{P} (Y > c) \text{ for all $c \in \R$}.
    $$
We start with the case where  $\widehat{\lambda}_2- \frac{k}{n} > 0$ showing
$$
  p^* \leq \mathbb{P} \left( \frac{n \, \widehat{\V}_b^2}{\widehat{\sigma}_Y^2} \geq k \right) = \P\left(\frac{(\widehat{\lambda}_1- \frac{k}{n})Q_1^2+(\widehat{\lambda}_2- \frac{k}{n})Q_2^2}{\frac{1}{n}(Q_3^2+\cdots+Q_{n-1}^2)} \geq k \right),
    $$ 
separately for the tree terms over which the minimum is taken.
For the first term, we need to show that $V \leq_{st} U$, where
    $$
U = \frac{(\widehat{\lambda}_1- \frac{k}{n})Q_1^2+(\widehat{\lambda}_2- \frac{k}{n})Q_2^2}{\frac{1}{n-3}(Q_3^2+\cdots+Q_{n-1}^2)}, \quad V= \left(\widehat{\lambda}_1- \frac{k}{n} \right)Q_1^2+ \left(\widehat{\lambda}_2- \frac{k}{n} \right)Q_2^2.
    $$
Also define,
$$
W = \frac{(\widehat{\lambda}_1- \frac{k}{n})Q_1^2+(\widehat{\lambda}_2- \frac{k}{n})Q_2^2}{\frac{1}{m-3}(Q_3^2+\cdots+X_{m-1}^2)},
$$
for some $m > n$ (where all $X_{j}^2$ are chisquare distributed random variables).

Let $H_U$, $H_V$ and $H_W$ denote the cumulative distribution functions of the random variables $U$, $V$ and $W$, respectively and let $h_U$, $h_V$ and $h_w$ denote their corresponding densities. Using the series representation in \cite[Equation (97)]{Kotz}, it follows that the family $\{H_V(ax), a > 0\}$ satisfies the monotone likelihood ratio property. Applying \cite[Proposition 2]{Rivest} now yields that $W$ is smaller than $U$ in the star-shaped order (cf. also Example 1 in \cite{Rivest}).

Using \cite[Theorem 1]{Dunkl} it is straightforward to show that
$$
h_U(0) \leq h_W(0).
$$
Applying \cite[Theorem 3]{Jeon} shows that  
$
W \leq_{st} U
    $
from which $V \leq_{st} U$ follows with a simple limit argument.

For the second term, we first observe that
$$
\frac{(\widehat{\lambda}_1- \frac{k}{n})Q_1^2}{\frac{1}{n-3}(Q_3^2+\cdots+Q_{n-1}^2)} \leq \frac{(\widehat{\lambda}_1- \frac{k}{n})Q_1^2+(\widehat{\lambda}_2- \frac{k}{n})Q_2^2}{\frac{1}{n-3}(Q_3^2+\cdots+Q_{n-1}^2)} 
$$
and hence
$$\mathbb{P} \left( \frac{n \, \widehat{\V}_b^2}{\widehat{\sigma}_Y^2} \leq k \right) \leq G_{F(1,n-3)} \left(\frac{k (n-3)}{\widehat{\lambda}_1 n - k} \right).
$$
The inequality for the third term is a direct consequence of \cite[Equation (32)]{Dunkl}.

For $p^{**}$, define the random variable
$$
Q = \frac{(\widehat{\lambda}_1 + \widehat{\lambda}_2- \frac{ 2\, k}{n})Q_1^2}{\frac{1}{n-2}(Q_3^2+\cdots+Q_{n-1}^2)},
$$

By \cite{Extremal}, the denominators of $Q$ and $U$ satisfy

$$
    \P\left( \Big( \big(\widehat{\lambda}_1- \frac{k}{n}\big)Q_1^2+\big(\widehat{\lambda}_2- \frac{k}{n}\big)Q_2^2\Big) \geq x\right) \leq \P\left(\Big(\widehat{\lambda}_1 + \widehat{\lambda}_2- \frac{ 2\, k}{n}\Big) Q_1^2 \geq x\right),
$$
whenever one of the expressions is smaller than $0.215$. It follows by a simple combinatorial argument that, for all $x$
$$
    \P (U \geq x) \leq \frac{1}{0.215} \P (Q \geq x),
$$
and $\frac{1}{0.215} < 5$.

Finally consider the case $\widehat{\lambda}_2- \frac{ k}{n} \leq 0$. Then
\begin{align*}
\Big(\widehat{\lambda}_1- \frac{k}{n} \Big) \, Q_1^2  - \frac{k}{n} \, Q_2^2 - \frac{k}{n} Q_3^2 - &\cdots - \frac{k}{n} Q_{n-1}^2 \leq T_n \\ \leq & \Big(\widehat{\lambda}_1- \frac{k}{n} \Big) \, Q_1^2   - \frac{k}{n} Q_3^2 - \cdots - \frac{k}{n} Q_{n-1}^2.
\end{align*}
The proof now follows by elementary transformations.
\qed

\bigskip

\Pro of Proposition 4.2. The kernel $k_b$ follows directly from using Equation (4.1) in the main manuscript with $x_0 = 1$. 
\qed

\bigskip

\Pro of Proposition 4.3. By direct evaluation, we see that the feature map $\boldsymbol{\Psi}(x) = (\psi_1(x), \psi_2(x))$ 
    $$
\psi_1(x) = \sqrt{\frac{b}{2}} (-1_{\{x=0\}} + 1_{\{x=2\}}) , \quad \psi_2(x) = \sqrt{\frac{4-b}{2}}(1_{\{x=1\}} - 1)
$$
satisfies
    $$
    k_b(x,x') = \langle \boldsymbol{\Psi}(x), \boldsymbol{\Psi}(x') \rangle.
    $$
It is easy to see that any translation of a feature map of $d_b$ is a feature map for $d_b$, which completes the proof.
\qed

\bigskip

\Pro of Theorem 4.4. For $t \in \{0,1,2\}$, let $V(t) = \sum_{j=1}^r B_j \phi_j (t)$. Since $\boldsymbol{\phi}(\cdot)$ is a feature map of $k_b$, we obtain 
    $$
   \E[V(s) V(t)]  = \sum_{j=1}^r \E [B_j^2] \, \phi_j(s) \, \phi_j(t) = k_b (s,t).
    $$
The Theorem now follows from Theorem 3 in \cite{DJ}.

\qed.

\bigskip

\Pro of Corollary 4.6. For $j \in \{1,\ldots,r\}$, define $D_j = A \, 1_{\{U=j\}}/c_j$. Then $\E[D_j] = 0$, 

$$
\E[D_j^2] = \frac{\E[A^2]  P(U=j)}{c_j^2} = \frac{\E[A^2]}{\sum_{k=1}^n c_k^2}.
$$
and, for $i \neq j$

$$
\E[D_i D_j] = \frac{\E[A^2]  P(U=i, U = j)}{c_i c_j} = 0.
$$

The result now follows from applying Theorem 4.4 with $B_j = D_j / \sqrt{\frac{\E[A^2]}{\sum_{k=1}^n c_k^2}}$ and $\tau$ replaced by $\tau \, \sqrt{\frac{\E[A^2]}{\sum_{k=1}^n c_k^2}}$.

\qed

\bigskip

\Pro of Theorem 4.7.

Define the stochastic process $V: \{0,1,2\} \to \R$ by,
    $$
    V(0) = 0, \quad V(1) =  B_1, \quad V(2) = (B_1 + B_2).
    $$
Then $\E[V(t)] = 0$, $\E[V(0)^2] = \E[V(0) V(1)] = \E[V(0) V(2)] = 0$, $\E[V(1)^2] = 1$,
    $$
\E[V(1) V(2)] = \E[B_1^2] + \E[B_1 B_2] = \frac{b}{2},
    $$
and
    $$
\E[V(2)^2 ] = \E[B_1^2] + 2\, \E[B_1 B_2] + E[B_2^2] = b.
    $$
Moreover, by choosing $x_0 = 0$ in Equation (4.1) of the main manuscript, we see that an alternative kernel induced by $d_b$ is $\widetilde{k}_b$ with
\begin{align*}
&\widetilde{k}_b(0,0) = \widetilde{k}_b(0,1) = \widetilde{k}_b(0,2) = 0, \\ &\widetilde{k}_b(1,1) = 2, \quad\widetilde{k}_b(1,2) = b , \quad \widetilde{k}_b(2,2) = 2b.
\end{align*}
Hence $E[V(s) V(t)] = \frac{1}{2} \widetilde{k}_b(s,t)$.
The result now follows by applying Theorem 3 in \cite{DJ}.
\qed

\bigskip

\Pro of Corollary 4.8.  
 While the proof is more instructive by constructing gamma-distributed variables and using Theorem 4.7, it leads to some  technicalities. To avoid these, we prove Corollary 4.8 directly, defining
    $$
V(0) = 0 , \quad V(1) = H A \quad V(2) = A.
    $$
It is easy to see that $V(0,0) = V(0,1) = V(0,2) = 0$. Moreover, by inserting the known first and second moments of the beta distribution, we obtain,
    $$
    \E[ V(1) V(1)] = \frac{1}{b} = \frac{1}{2 b } \widetilde{k}_b(1,1),
    $$
    $$
 \E[ V(1) V(2)] = \frac{1}{2} = \frac{1}{2 b } \widetilde{k}_b(1,2),
    $$
    $$
   \E[ V(2) V(2)] = 1 = \frac{1}{2 b } \widetilde{k}_b(2,2),
    $$
where $\widetilde{k}_b$ is defined in the proof of Theorem 4.6.
Applying Theorem 3 in \cite{DJ} completes the proof.
\qed
  
\bigskip


\Pro of Theorem 5.3. We will use the same notation as in the proof of Theorem 3.3.
   Moreover, let $H_Z$ denote the projection matrix  
    $$
        H_Z =  {\bZ} ({\bZ}^t {\bZ})^{-1} {\bZ}^t,
    $$
It is straightforward to see that $\widehat{\V}_b^2(\bX,\bY; \bZ)$ can be written as
    \begin{align*}
n \, \V_b^2(\bX,\bY; \bZ) &= \frac{1}{n} \bY^t (I-H_Z) \bU \bU^t (I-H_Z) \bY \\ &= \bv^t \bv,
\end{align*}
with 
$$
\bv = \frac{1}{\sqrt{n}} \bU^t (I-H_Z) \bY.
$$
Since $(I-H_Z) {\bZ} = 0$ , $\bv$ can alternatively be written as,
\begin{align}
\bv &= \frac{1}{\sqrt{n}} (\bU- {\bZ} \alpha)^t (I-H_Z) (\bY- {\bZ} \gamma) \\
&= \frac{1}{\sqrt{n}} (\bU- {\bZ} \alpha)^t (\bY- {\bZ} \gamma) - \frac{1}{\sqrt{n}} (\bU- {\bZ} \alpha)^t H_Z (\bY-{\bZ} \gamma) \label{eq:vdecomp2} ,
\end{align}
where we denote $\alpha=\E[Z Z^t]^{-1} E[Z U]$
We first consider the second term in Equation \eqref{eq:vdecomp2},
\begin{align*}
   \frac{1}{\sqrt{n}} (\bU- {\bZ} \alpha)^t H_Z (\bY- {\bZ} \gamma) &= 
    \frac{1}{n^{3/2}} (\bU- {\bZ} \alpha)^t {\bZ} \,  (n^{-1} {\bZ}^t {\bZ})^{-1} {\bZ}^t (\bY- {\bZ} \gamma). \\
\end{align*}
With similar arguments as in the proof of Theorem 3.3, it follows that $vec((\bU- {\bZ} \alpha)^t {\bZ})$ and ${\bZ}^t (\bY- {\bZ} \gamma)$ converge to normal distributions with mean $0$, whereas $n^{-1} {\bZ}^t {\bZ}$ converges to the (augmented) covariance matrix of $Z$. Hence, this term converges to $0$.

We now consider the first term in Equation \eqref{eq:vdecomp}. Applying the multivariate CLT and using that $U$ and $Y$ are independent, we observe that, for $n \to \infty$
$$
\frac{1}{\sqrt{n}} (\bU- {\bZ} \alpha)^t (\bY- {\bZ} \gamma)  \stackrel{\mathcal{D}}{\longrightarrow} N(0,\Gamma),
$$
where 
    $$
\Gamma =\sigma_\varepsilon^2 K 
    $$

The rest of the proof is analogous to the proof of Theorem 3.3.
\qed

\bigskip

\Pro of Theorem 5.4. We use the same notation as in the proofs to Theorem 3.3, Theorem 3.4 and Theorem 5.3. As in the proof of Theorem 5.3, we write
$\widehat{V}_b^2(\bX,\bY ;\bZ)$ as
    \begin{align*}
n \, \V_b^2(\bX,\bY) &= \frac{1}{n} \bY^t (I_n-H_Z) \bU \bU^t (I_n-H_Z) \bY \\ &= \frac{1}{n}  (\bY-{\bZ} \gamma)^t (I_n-H_Z) \bU \bU^t (I_n-H_Z) (\bY- {\bZ} \gamma).
\end{align*}
Consequently,
    $$
 \frac{n \, \widehat{\V}_b^2(\bX,\bY;\bZ)}{\widehat{\sigma}_\varepsilon^2}  =  \frac{(\bY-{\bZ} \gamma)^t (I_n-H_Z) \bU \bU^t (I_n-H_Z) (\bY- {\bZ} \gamma)}{(\bY-{\bZ} \gamma)^t (I_n-H_Z) I_n (I_n-H_Z) (\bY- {\bZ} \gamma)}.
    $$
Hence
$$
 \left\{ \frac{n \, \widehat{\V}_b^2(\bX,\bY;\bZ)}{\widehat{\sigma}_\varepsilon^2} \geq k \right\}
$$
is obviously equivalent to
$$
 \{ (\bY-{\bZ} \gamma)^t (I_n-H_Z) \frac{1}{n} (\bU \bU^t - k I_n) (I_n-H_Z)(\bY-{\bZ} \gamma)\geq 0\}.
$$
Now consider the matrix
\begin{align*}
 \frac{1}{n} (I_n-H_Z) (\bU \bU^t  - k I_n) (I_n-H_Z) =  \frac{1}{n} (I_n-H_Z) \bU \bU^t   (I_n-H_Z) - \frac{k}{n} \, (I_n - H_Z).
 \end{align*}
 Let $\bo_{n-p-1},\ldots \bo_n$  denote orthogonal eigenvectors $(I_n - H_Z)$ corresponding to eigenvalue $0$. Then $\bo_{n-p-1},\ldots \bo_n$  are obviously also eigenvectors to eigenvalue $0$ of the matrix $(I_n-H_Z) \bU \bU^t   (I_n-H)$. Augmenting $\bo_{n-p-1},\ldots \bo_n$ to an orthogonal basis (represented by the matrix $O$) of $(I_n-H_Z) \bU \bU^t   (I_n-H_Z)$, we obtain,
 \begin{align*}
\frac{1}{n}  (I_n-H_Z) \bU \bU^t   (I_n-H_Z) - \frac{k}{n} \, (I_n - H_Z) 
 &= O \widehat{\Lambda} O^t -  O  D_{n-p-1} O^t,
\end{align*}
where $D_{n-p-1}$ is a diagonal matrix where the entries of the diagonal are $p+1$ zeros and $n-p-1$ times the value $k/n$ and $\widehat{\Lambda}$ is a diagonal matrix with diagonal $(\widehat{\lambda_1},\widehat{\lambda_2},0,\ldots,0)$. Since the standard normal distribution is invariant under orthogonal transformations we obtain that
\begin{align*}
&(\bY-{\bZ} \gamma)^t (I_n-H_Z) \frac{1}{n} (\bU \bU^t - k I_n) (I_n-H_Z)(\bY-{\bZ} \gamma) \\ &\stackrel{\mathcal{D}}{=} (\widehat{\lambda}_1 - k/n) Q_1^2 + (\widehat{\lambda}_2 - k/n) Q_2^2 - k/n Q_3^2 - \cdots - k/n Q_{n-p-1}^2.
\end{align*}
\qed

\section{Additional results}

\subsection{Comments on Theorem 4.7 and an extension of Corollary 4.8}

For constructing bivariate random vectors with zero mean for which the marginals have equal or opposite sign, we note that for any pair of non-negative random variables $G =(G_1,G_2)^t$ we can use a random variable $A$ with $P(A=1) = P(A=-1) = \tfrac12$, independent of $G$ to construct mean zero random variables
    $$
  B = \begin{pmatrix} B_1 \\ B_2 \end{pmatrix} = \begin{pmatrix} A G_1 \\ A G_2 \end{pmatrix}, \quad 
    \widetilde{B} = \begin{pmatrix} \widetilde{B}_1 \\ \widetilde{B}_2 \end{pmatrix} = \begin{pmatrix} A G_1 \\ - A G_2 \end{pmatrix}.
    $$
Then the marginals of $B$ are have equal sign, the ones of $\widetilde{B}$ have opposing signs and 
$$
    Cor(B_1,B_2) = \frac{\E[A_1 A_2]}{\sqrt{\E[A_1^2] \E[A_2^2]}} \quad \quad Cor(\widetilde{B_1},\widetilde{B_2}) = -\frac{- \E[A_1 A_2]}{\sqrt{\E[A_1^2] \E[A_2^2]}}
$$
Choosing $G_1$ and $G_2$  gamma-distributed with equal rate parameter and shape parameter $\frac{a-2}{4-a}$ ($a \in (2,4)$) leads to (cf. Corollary 4.8),
    $$
Cor(B_1,B_2) = \frac{a}{2}-1
    $$
and consequently,
    $$
Cor(\widetilde{B_1},\widetilde{B_2}) = 1- \frac{a}{2}.
    $$
Now let $b \in (0,2)$ and choose $a= 4-b$. Then
    $$
Cor(\widetilde{B_1},\widetilde{B_2}) = \frac{b}{2} - 1.
    $$
One directly obtains the following Corollary of Theorem 4.7, which - in the light of the above can be regarded as an analogue of Corollary 4.8 for $b \in (0,2)$.

\begin{corollary} \label{cor:beta}
	Consider the distance $d_b$ with $b \in (0,2)$ and
	assume the model
	\begin{align*}
	Y_i = \begin{cases}  \mu_Y + \varepsilon, &\text{ if $x_i =0$}, \\
	\mu_Y + \tau G_1 A + \varepsilon,  &\text{ if $x_i =1$}\\
	 \mu_Y + \tau (G_1-G_2) A + \varepsilon	&\text{ if $x_i =2$,}
	\end{cases}
	\end{align*}
	where $\mu_Y$ is known, $\tau \in \R$, $\varepsilon \sim \mathcal{N}(0,\sigma^2)$ and $G =(G_1, G_2)^t$ are independently gamma-distributed with shape parameters  $\frac{2-b}{b}$ and equal rate parameters; $A$ is a random variable, independent of $G$ with $\E[A]=0$ and $\E[A^2]=1$ (e.g. $P(A = 1) = P(A = - 1) = \frac{1}{2}$).
 Then the locally most powerful test for testing 
    $
        H_0 : \tau^2 = 0 \text{ against } H_1 : \tau^2 > 0
    $ is given by Equation (4.5) in the main manuscript.
\end{corollary}

Hence, for $b \in (0,2)$, $\V_b$ can be interpreted as the locally most powerful test in the case where the heterozygous effect is distributed as $\frac{G_1}{G_1 - G_2}$, where $G_1,G_2$ are independently gamma-distributed with shape parameters  $\frac{2-b}{b}$$.